\documentclass[journal ]{new-aiaa}
\usepackage[utf8]{inputenc}
\usepackage{textcomp}

\usepackage{graphicx}
\usepackage{amsmath}
\usepackage[version=4]{mhchem}
\usepackage{siunitx}
\usepackage{longtable,tabularx}
\usepackage{comment}
\title{Terminal Soft Landing Guidance Law using Analytic Gravity Turn Trajectory}

\author{Seungyeop Han \footnote{Ph.D. Student, Daniel Guggenheim School of Aerospace Engineering}}
\affil{Georgia Institute of Technology, Atlanta, Georgia, 30332}
\author{Byeong-Un Jo \footnote{Assistant Professor, Department of Aerospace Engineering (Former Postdoctoral Fellow in Daniel Guggenheim School of Aerospace Engineering at Georgia Institute of Technology)}}
\affil{Sejong University, Seoul 05006, Republic of Korea}
\author{Koki Ho \footnote{Dutton-Ducoffe Professor, Associate Professor, Daniel Guggenheim School of Aerospace Engineering, AIAA Senior Member, kokiho@gatech.edu (Corresponding Author)}}
\affil{Georgia Institute of Technology, Atlanta, Georgia, 30332}

\begin{document}
\footnotetext{This paper is a substantially revised version of Paper AAS 23-152 presented at 33rd AAS/AIAA Space Flight Mechanics Meeting, Austin, TX, January 15-19, 2023, with new mathematical analysis and results.}
\maketitle

\begin{abstract}
This paper presents an innovative terminal landing guidance law that utilizes an analytic solution derived from the gravity turn trajectory. The characteristics of the derived solution are thoroughly investigated, and the solution is employed to generate a reference velocity vector that satisfies terminal landing conditions. A nonlinear control law is applied to effectively track the reference velocity vector within a finite time, and its robustness against disturbances is studied. Furthermore, the guidance law is expanded to incorporate ground collision avoidance by considering the shape of the gravity turn trajectory. The proposed method's fuel efficiency, robustness, and practicality are demonstrated through comprehensive numerical simulations, and its performance is compared with existing methods.
\end{abstract}

\section*{Nomenclature}

{\renewcommand\arraystretch{1.0}
\noindent\begin{longtable*}{@{}l @{\quad=\quad} l@{}}
$\textbf{\textit{a}}$  & geometric vector \\
$\hat{\textbf{\textit{a}}}$ & unit vector of $\textbf{\textit{a}}$ \\
$\textit{\textit{a}}$ & magnitude of $\textbf{\textit{a}}$ or scalar value \\
$\textbf{\textit{a}}^A$ & $\textbf{\textit{a}}$ expressed in frame $A$ \\
$^{A} \dot{\textbf{\textit{a}}}$ & time derivative of $\textbf{\textit{a}}$ with respect to frame $A$ \\
$^{A} \dot{\textbf{\textit{a}}}^A$ & $^{A} \dot{\textbf{\textit{a}}}$ expressed in frame $A$ \\
\end{longtable*}}

\section{Introduction}
\lettrine{M}{o}on and Mars explorations have captivated researchers for an extended period, leading to numerous projects in these domains. Notable examples include the Apollo and Artemis missions for lunar exploration, as well as the Opportunity, Curiosity, and Perseverance missions for Mars exploration \cite{Moon_landing, Mars_landing}. These projects require the soft landing of sophisticated landers or rovers to facilitate their exploration mission, but previous missions have experienced landing errors of several kilometers \cite{Braun_2007, Steinfeldt_2010}. Consequently, extensive research has been conducted on powered descent and pinpoint landing methods, utilizing thrusters. Additionally, there has been a focus on small celestial body landing missions that do not have atmospheric interference \cite{Kikuchi_2020}, as well as investigations into the landing techniques of lifting bodies, leveraging their aerodynamic properties \cite{Sostaric_2017, Tahk_2016}. In all cases, the development of a suitable terminal landing guidance law is crucial to achieve a safe landing at the intended site.

Recent works related to the problem can be broadly categorized into two groups: computational guidance and analytical feedback guidance, and the former approaches have gained popularity due to their optimality and ability to handle constraints effectively.
One notable approach is the use of convex programming with lossless convexification to solve the fuel optimal problem \cite{Acikmese_2007, blackmore2010minimum}. Subsequent research has shown that fuel optimal trajectories can be obtained within limited computation time \cite{Dueri_2017, scharf2017implementation}. In Ref.~\cite{mao2016successive}, the successive convexification algorithm was introduced to solve general non-convex optimal control problems. It is applied to the landing problem, including nonconvex aerodynamics \cite{liu2019fuel}, thrust and curvature constraints \cite{CUI2022313}, 6 degrees of freedom with free-final time \cite{szmuk2018successive}, and state-triggered constraints \cite{szmuk2020successive, reynolds2020dual}. Yet another famous approach for solving the fuel optimal problem is using optimal control theory \cite{Lu_2018}. This method converts the landing problem into a multivariable root-finding problem via optimality conditions.
 In Ref.~\cite{reynolds2020optimal}, an optimal powered descent landing under planar motion with independent torque dynamics is investigated, and a numerically efficient algorithm is introduced \cite{spada2023direct}. Additionally, new approaches have emerged utilizing techniques like supervised training with deep neural networks \cite{sanchez2018real} and reinforcement learning \cite{gaudet2014adaptive} to tackle the problem. Linear programming based on the gravity turn trajectory has also been employed \cite{yang2022gravity}.

In some practical systems, classical feedback guidance algorithms are preferred due to their general advantages in handling unknown disturbances and computational efficiency. The computational efficiency of these algorithms enables fast command updates and allows for the integration of other computationally heavy algorithms, such as hazard detection and navigation, which play a crucial role in achieving precise and safe landing. Polynomial guidance is one well-known approach to feedback-type guidance methods due to its use in the Apollo mission \cite{KLUMPP1974133}. The polynomial guidance concept has been revisited in simpler \cite{Wong_2013} and generalized forms \cite{Lu_APDG, Lu_fractional}. These works demonstrate that polynomial guidance can achieve various landing constraints with efficient fuel usage, given appropriate powered descent initiation conditions. In contrast, some works utilize the sliding mode control (SMC) to address the problem \cite{Furfaro_smc, YANG_smc, ZHANG_smc}, which generally offers robustness against disturbances due to its control structure.

Another well-known guidance concept is a variant of energy-optimal guidance \cite{Souza_opt} called zero-effort-miss and zero-effort-velocity (ZEM/ZEV) guidance \cite{EBRAHIMI2008556}. This concept has gained popularity in research and applied to various applications due to its simplicity and high performance \cite{Guo_app}.  In Ref. \cite{WIBBEN_zemzev}, a SMC structure is incorporated with ZEM/ZEV guidance to enhance its performance, while \cite{ZHOU_zemzev} presents an improved version of the original ZEM/ZEV guidance that addresses the chattering problem commonly associated with SMC. Also, improved ZEM/ZEV guidance with the aid of machine learning is introduced \cite{furfaro2020adaptive} and the optimal cost function is modified to include collision avoidance capability \cite{ZHANG_zemzev}. Furthermore, \cite{CUI_zemzev} introduces a constant-thrust phase alongside the ZEM/ZEV guidance law to shape the trajectory's curvature. On the other hand, to address a wide range of constraints such as ground surfaces and curvature, a hybrid computational guidance approach that combines online optimization and trajectory prediction methods has been employed. For instance, in \cite{Guo_waypoint}, a multiple-waypoint ZEM/ZEV method is proposed, which achieves a fuel-efficient trajectory while satisfying state constraints. In Ref. \cite{ZHANG_zemopt}, a numerical integration method with parameter optimization is utilized to meet multiple constraints. Furthermore, a two-phase ZEM/ZEV control logic is employed, incorporating numerical integration and parameter optimization techniques to achieve a fuel-efficient trajectory with ground collision avoidance \cite{Wang_twophase}.

The gravity turn trajectory (GT) is a classic and practical approach for terminal landing, which has been successfully employed in multiple previous missions \cite{cheng1966design, Desai_2011}. Previous works have published guidance laws based on the GT \cite{cheng1966design, CITRON_1964}, but they are limited to handling the soft landing of two-dimensional planar motion. Subsequent studies extended the approach by incorporating nonlinear control theory \cite{McInnes_1999} and exploring the logic for three-dimensional motion \cite{Chomel_2009}. The GT approach offers advantages such as simple thrust control, concave curvature trajectory, vertical landing capability, analytic formulation, and fuel optimality for soft landing. However, it does not inherently achieve pinpoint landing, which is a critical requirement. This paper recognizes the promising characteristics of the GT and utilizes the GT solution to design a dedicated pinpoint landing guidance law for the terminal landing phase.

The approach presented in this paper combines the strengths of feedback guidance and the GT. The resulting trajectory exhibits a concave down shape relative to the landing site, which facilitates obstacle avoidance and ensures availability of sensor field of view. Furthermore, the thrust vector aligns with the local gravitational direction at the final moment, ensuring the correct vehicle attitude. Unlike many recent approaches, the proposed method is computationally efficient as it does not require online optimization, trajectory propagation, and offline time-to-go estimation. The trajectory is fuel-efficient and approaches near-optimal performance under specific analytically characterizable conditions. Lastly, the proposed guidance law guarantees finite-time error convergence and robustness against disturbances.

The remainder of the paper is structured as follows. Section \ref{sec2} provides a review of the analytic solution of the gravity turn trajectory and discusses its characteristics. The method for generating a desired velocity vector that guides a lander to the landing site is explained in Section \ref{sec3}. In Section \ref{sec4}, a nonlinear control law for tracking the velocity vector, developed in the previous section, is proposed. The stability and robustness of the control law are also proven. Furthermore, Section \ref{sec4} presents the ground collision avoidance logic and the thrust command distribution logic. The effectiveness of the proposed method is demonstrated through numerical simulations under various scenarios, and the performance of the proposed method is also compared with existing methods in Section \ref{sec5}.

\section{Analytic Solution of Planar Gravity Turn Trajectory} \label{sec2}
The terminal phase, which is the final phase of Entry, Descent, and Landing (EDL), exhibits distinct dynamic characteristics compared to other landing phases. In the terminal landing phase, both the altitude and velocity of the lander are relatively low. As a result, the Coriolis force can be ignored, and the flat planet model can be reasonably assumed, allowing the analytic gravity-turn solution. The objectives of this section are to obtain the analytic solution for the gravity turn trajectory and to investigate its useful properties for designing a guidance law. 
We emphasize that the solution is to generate a reference trajectory to guide a lander toward a landing site and it is not an actual flight trajectory.

\subsection{Analytic Solution of Powered Descent Gravity Turn Trajectory}
The gravity turn trajectory refers to a trajectory where the gravitational force plays a major force in altering the flight-path angle, and this can be achieved by aligning the thrust vector with the velocity vector. Under the assumptions of a constant mass and negligible disturbances, the planar gravity turn motion of a vehicle can be expressed as follows:
\begin{subequations}\label{eom_planar}
\begin{equation} \label{eom_planar1}
m\frac{dv}{dt} = -T - mg\sin\gamma
\end{equation}
\begin{equation} \label{eom_planar2}
mv\frac{d\gamma}{dt} = -mg\cos\gamma
\end{equation}
\begin{equation} \label{eom_planar3}
\frac{dx}{dt} = v\cos\gamma
\end{equation}
\begin{equation} \label{eom_planar4}
\frac{dz}{dt} = v\sin\gamma
\end{equation}
\end{subequations}
\noindent where $T$ is the thrust, $m$ is the mass, $g$ is the local gravitational acceleration, $v$ is the speed, $\gamma$ is the flight-path angle, $x$ is the downrange and $z$ is the height of the lander along the planar motion. 
Along the trajectory where $v > 0$ and $\gamma \in (-\frac{\pi}{2}, \frac{\pi}{2})$, there is a monotonic relationship between $\gamma$ and $t$ since the time derivative of $\gamma$ becomes strictly negative. This relationship allows us to change the independent variable from $t$ to $\gamma$. This can be achieved by dividing Eqs.~\eqref{eom_planar1}, \eqref{eom_planar3}, and, \eqref{eom_planar4} by Eq.~\eqref{eom_planar2}, yielding the following expressions:
\begin{subequations} \label{mod_grav_eom}
\begin{equation} \label{mod_grav_eom_1}
\frac{dv}{d\gamma} = v\tan\gamma + \beta v \sec\gamma
\end{equation}
\begin{equation} \label{mod_grav_eom_2}
\frac{dt}{d\gamma} = -\frac{v}{g}\sec\gamma
\end{equation}
\begin{equation} \label{mod_grav_eom_3}
\frac{dx}{d\gamma} = -\frac{v^2}{g}
\end{equation}
\begin{equation} \label{mod_grav_eom_4}
\frac{dz}{d\gamma} = -\frac{v^2}{g}\tan\gamma
\end{equation}
\end{subequations}
 where $\beta \equiv T/mg$ is the constant normalized acceleration (thrust to weight ratio) by thruster. Then the analytic solution for Eqs.~\eqref{mod_grav_eom} for $\gamma_0 \in (-\pi/2, \pi/2)$ is known as follows \cite{HAN2016_gt}:
\begin{subequations} \label{anal_sol}
\begin{equation} \label{anal_sol_1}
v\left(  \gamma \right) = C \sec\gamma \left( \sec\gamma + \tan\gamma \right)^\beta
\end{equation}
\begin{equation} \label{anal_sol_2}
t\left(  \gamma \right) = t_0 - \frac{C}{g}\left[F_t(\gamma) - F_t(\gamma_0) \right]
\end{equation}
\begin{equation} \label{anal_sol_3}
x\left(  \gamma \right) = x_0 - \frac{C^2}{g}\left[F_x(\gamma) - F_x(\gamma_0) \right]
\end{equation}
\begin{equation} \label{anal_sol_4}
z\left(  \gamma \right) = z_0 - \frac{C^2}{g}\left[F_z(\gamma) - F_z(\gamma_0) \right]
\end{equation}
\end{subequations}
where subscript $0$ implies the initial states, $C$ is the integration constant defined as:
\begin{equation} \label{anal_int_const}
C =  \frac{ v_0 }{ \sec\gamma_0 \left( \sec\gamma_0 + \tan\gamma_0 \right)^\beta}
\end{equation}
$F_t$, $F_x$, and $F_z$ represent the indefinite integral computed for $\beta >1$ as:
\begin{subequations} \label{anal_int}
\begin{equation} \label{anal_int_1}
F_t\left(  \gamma \right) = \frac{1}{\beta^2-1}(\beta\sec\gamma - \tan\gamma)(\sec\gamma + \tan\gamma)^\beta
\end{equation}
\begin{equation} \label{anal_int_2}
F_x\left(  \gamma \right) = \frac{1}{4\beta^2-1}(2\beta\sec\gamma - \tan\gamma)(\sec\gamma + \tan\gamma)^{2\beta}
\end{equation}
\begin{equation} \label{anal_int_3}
F_z\left(  \gamma \right) = \frac{1}{4\beta^2-4}(2\beta\sec\gamma\tan\gamma - 2\tan^2\gamma - 1)(\sec\gamma + \tan\gamma)^{2\beta}
\end{equation}
\end{subequations}
The modified equations should be used when $\beta \leq 1$ which is not a case of interest for the problem \cite{HAN2016_gt}. 

\subsection{Characteristics of Powered Descent Gravity-Turn Trajectory} 
Before applying the solution for guidance law design, it is important to understand the properties of the solution trajectory. Without loss of generality, a flight path angle is assumed to be $\gamma \in [-\frac{\pi}{2}, \frac{\pi}{2})$ only for analysis purposes.

\textbf{\textit{Proposition 1:}} Along the trajectory, the flight-path angle $\gamma$ asymptotically converges to $-\frac{\pi}{2}$, resulting in the thrust direction becoming parallel to the nadir direction.

\noindent
\textbf{\textit{Proof: }} Let $V \equiv (\gamma + \frac{\pi}{2})^2$, then $V > 0$ for $\gamma \neq -\frac{\pi}{2}$ and $V = 0$ for $\gamma = -\frac{\pi}{2}$. The time derivative of $V$ along the trajectory is:
\begin{equation} \label{anal_prop1}
\dot{V} = -2 \left(\gamma + \frac{\pi}{2} \right) \frac{g}{v} \cos \gamma  < 0
\end{equation}
and $v>0$ for $\gamma \in (-\frac{\pi}{2}, \frac{\pi}{2})$. By the Lyaponov stability theorem, $V \to 0$ or equivalently $\gamma \to -\frac{\pi}{2}$ as $t \to \infty$. Therefore, the lower limit value of $\gamma$ of gravity-turn is $-\frac{\pi}{2}$, indicating that the velocity and thrust direction become parallel to the nadir during descent.

\textbf{\textit{Proposition 2:}} Along the trajectory with $\beta > 1$, $v \to 0$ as $\gamma \to -\frac{\pi}{2}^+$ within finite time. Consequently, downrange as well as altitude variation are finite.

\noindent
\textbf{\textit{Proof: }} For $a\geq 0, b>0$, following identity holds:
\begin{equation} \label{anal_prop2_1}
\begin{aligned}
\sec^a\gamma (\sec\gamma+\tan\gamma)^b 
&= \frac{(1+\sin\gamma)^b(1-\sin\gamma)^b}{\cos^{a+b}\gamma(1-\sin\gamma)^b} = \frac{\cos^{b-a}\gamma}{(1-\sin\gamma)^b} \\
\end{aligned}
\end{equation}

\noindent
One can readily show that limit value of Eq.~\eqref{anal_prop2_1} when $b-a>0$ is:
\begin{equation} \label{anal_prop2_2}
\lim_{\gamma \to -\frac{\pi}{2}^+} \frac{\cos^{b-a}\gamma}{(1-\sin\gamma)^b} = 0 \quad \text{for} \quad b-a>0
\end{equation}

\noindent
The Eq.~\eqref{anal_sol_1} belongs to the particular case of Eq.~\eqref{anal_prop2_1} with $a=1$ and $b=\beta>1$ by assumption, hence $v \to 0$ as $\gamma \to -\frac{\pi}{2}^+$. Next, the absolute value of indefinite integral $F_t$, $F_x$, and $F_z$ can be bounded for $\gamma \in (-\frac{\pi}{2}, \frac{\pi}{2})$ as:
\begin{equation} \label{anal_prop2_3}
\begin{aligned}
  &0 \leq \vert F_t(\gamma) \vert \leq  \frac{1}{\beta^2-1 }(\beta\sec\gamma + \sec\gamma)(\sec\gamma + \tan\gamma)^\beta \\
  &0 \leq \vert F_x(\gamma) \vert \leq  \frac{1}{4\beta^2-1 }(2\beta\sec\gamma + \sec\gamma)(\sec\gamma + \tan\gamma)^{2\beta} \\
  &0 \leq \vert F_z(\gamma) \vert \leq  \frac{1}{4\beta^2-4 }(2\beta\sec^2\gamma + 2\sec^2\gamma + 1)(\sec\gamma + \tan\gamma)^{2\beta} \\
\end{aligned}
\end{equation}
from the conditions: $0 \leq \vert \tan\gamma \vert \leq \sec\gamma$, $0 \leq \vert \tan^2\gamma \vert \leq   \sec^2\gamma $, and $0 \leq \vert \sec\gamma\tan\gamma \vert \leq \sec^2\gamma $ for $\gamma \in (-\frac{\pi}{2}, \frac{\pi}{2})$. The upper bounds of all indefinite integrals satisfy the condition of Eq.~\eqref{anal_prop2_1} with $(a,b) = (1,\beta)$, $(1,2\beta)$ and $(2,2\beta)$, respectively. Therefore, the limit value of Eqs.~\eqref{anal_int} can be obtained by applying the squeeze theorem as:
\begin{equation} \label{anal_prop2_4}
\lim_{\gamma \to -\frac{\pi}{2}^+} F_t\left(  \gamma \right) 
= \lim_{\gamma \to -\frac{\pi}{2}^+}F_x\left(  \gamma \right) 
= \lim_{\gamma \to -\frac{\pi}{2}^+} F_z\left(  \gamma \right) = 0
\end{equation}

Propositions 1 and 2 imply that variation of time, downrange, and altitude are finite during the gravity turn phase, given that $\beta > 1$. In other words, $v \to 0$, $\gamma \to -\frac{\pi}{2}$ within finite time, and the terminal value of Eq.~\eqref{anal_sol} can be simplified as follow \cite{Han_MS_2017}:
\begin{subequations} \label{anal_intsol}
\begin{equation} \label{anal_intsol1}
x_f = x_0 + \frac{v^2_0}{(4\beta^2-1)g}(2\beta\cos\gamma_0 - \sin\gamma_0\cos\gamma_0)
\end{equation}
\begin{equation} \label{anal_intsol2}
z_f = z_0 + \frac{v^2_0}{(4\beta^2-4)g}(2\beta\sin\gamma_0 - \sin^2\gamma_0 - 1)
\end{equation}
\begin{equation} \label{anal_intsol3}
t_f = t_0 + \frac{v_0}{(\beta^2-1)g}(\beta - \sin\gamma_0)
\end{equation}
\end{subequations}

\noindent
where subscript $0$ implies the initial states and subscript $f$ means the final states (when $v_f = 0$), respectively. 

\section{Velocity Vector for Terminal Landing} \label{sec3}
Referring to Eq.~\eqref{anal_intsol1} and Eq.~\eqref{anal_intsol2}, and treating $x_f$ and $z_f$ as the desired landing location give the two nonlinear equations described by initial states ($x_0$, $z_0$, $v_0$, $\gamma_0$) and trajectory parameters ($\beta$, $g$). In the following analysis, the initial states will be replaced with the current states, and the subscript 0 will be omitted for readability. Consequently, the nonlinear state equations Eqs.~\eqref{anal_intsol1} and \eqref{anal_intsol2} can be re-expressed as follows:
\begin{subequations} \label{eqn_13}
\begin{equation}  \label{eqn_13a}
f_x(v, \gamma, x_\text{go}) = v^2(2\beta c_\gamma - s_\gamma c_\gamma) - (4\beta^2-1) g x_\text{go}
\end{equation}
\begin{equation} \label{eqn_13b}
f_z(v, \gamma, z_\text{go}) = v^2(2\beta s_\gamma - s_\gamma^2 - 1) - (4\beta^2-4) g z_\text{go}
\end{equation}
\end{subequations}
where $x_\text{go} \equiv x_f-x$, $z_\text{go} \equiv z_f-z$, $c_\gamma \equiv \cos\gamma$, and $s_\gamma \equiv \sin\gamma$. Note that there are infinitely many solutions sets $\{v, \gamma, x_\text{go}, z_\text{go}\}$ satisfying $f_x=f_z=0$, but it turns out that $v$ and $\gamma$ are uniquely determined for specific $x_\text{go}$, $z_\text{go}$, and $\beta > 1$. This relationship enables us to generate a velocity vector field, which will be elucidated in this section. In this section, $v$ and $\gamma$ satisfying Eqs.~\eqref{eqn_13} will be expressed as $v^\ast$ and $\gamma^\ast$ to indicate that they are the solution.

\subsection{Uniqueness and Existence of Velocity Vector} \label{sec3.1}
By proper selection of guidance reference frame, which will be explained in the subsequent section, it is always possible to make $x_\text{go} \geq 0$. Therefore, for the sake of simplifying the derivation process, the paper assumes that $x_\text{go} \geq 0$, or equivalently, $\gamma \in [-\frac{\pi}{2}, \frac{\pi}{2}]$.

\textbf{\textit{Case 1: $x_\text{go} \neq 0$}} \\
Rearranging $f_x(v,\gamma)=f_z(v,\gamma)=0$ defines single-variable function $h(\gamma)$ with $\kappa \equiv \frac{(4\beta^2-4)z_\text{go}}{(4\beta^2-1)x_\text{go}}$ as:
\begin{equation}  \label{h_eqn}
h(\gamma) \equiv \frac{2\beta s_\gamma - s^2_\gamma - 1}{2\beta c_\gamma - s_\gamma c_\gamma} - \kappa  \end{equation}

\noindent By the assumption $\kappa$ is bounded, and $h(\gamma) \to \infty$ as $\gamma \to \frac{\pi}{2}^-$ and $h(\gamma) \to - \infty$  as $\gamma \to -\frac{\pi}{2}^+$. On the other hand, derivative of $h(\gamma)$ with respect to $\gamma$ is:
\begin{equation}  \label{h_prime}
\frac{dh}{d\gamma}
= \frac{ 3(\beta - s_{\gamma})^2 + \beta^2 - 1
}{\left(2\beta c_{\gamma} - s_{\gamma}c_{\gamma} \right)^2}
\end{equation}

\noindent
and it turns out to be strictly positive since $\beta > 1$. The intermediate value theorem, combined with the strict monotonicity of $h(\gamma)$, guarantees the existence of a unique solution $\gamma^\ast$ that makes $h(\gamma^\ast)=0$. Due to the monotonicity of $h(\gamma)$, any single-variable numerical root-finding method can obtain an accurate solution within a few iterations. Note that the Newton's method works well with following initial guess:
\begin{equation} \label{eqn_16}
\gamma^{(k+1)} = \gamma^{(k)} - \frac{h\left(\gamma^{(k)}\right)}{\frac{dh}{d\gamma}\left(\gamma^{(k)}\right)} 
,\quad \gamma^{(0)} = \tan^{-1} \left(\frac{z_\text{go}}{x_\text{go}} \right)
\end{equation}
Once $\gamma^\ast$ is determined, then $v^\ast$ can be computed by original equation as follows
\begin{equation} \label{v_sol}
v^\ast = \sqrt[]{ \frac{(4\beta^2-1)g x_\text{go}}{ (2\beta - s_{\gamma^\ast}) c_{\gamma^\ast}}}
\end{equation}

\textbf{\textit{Case 2: $x_\text{go} = 0$}} \\
The condition $x_\text{go} = 0$ means $c_{\gamma^\ast} = 0$, and it implies that either $\gamma^\ast = \frac{\pi}{2}$ or $\gamma^\ast = -\frac{\pi}{2}$. The case in which the landing site is located above the lander is ignored, so the correct solution can be determined as follows
\begin{equation} \label{1d_sol}
(v^\ast, \gamma^\ast) = \left( \sqrt[]{2(\beta-1)g \lvert z_\text{go}} \rvert, -\frac{\pi}{2} \right)
\end{equation}

\subsection{Properties of Velocity Vector}
\begin{figure}[b!]
\centering
\includegraphics[width=.48\textwidth]{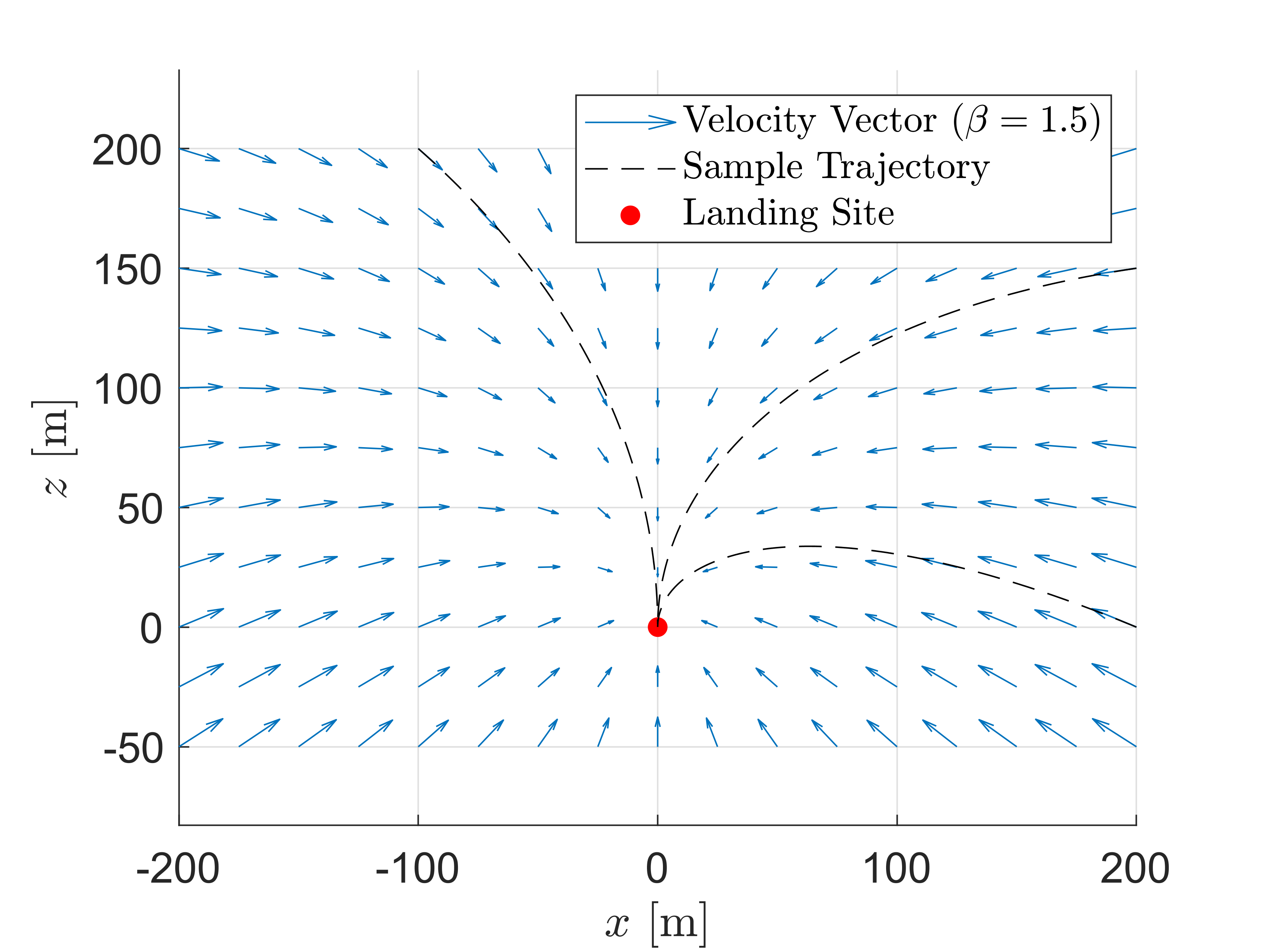} 
\includegraphics[width=.48\textwidth]{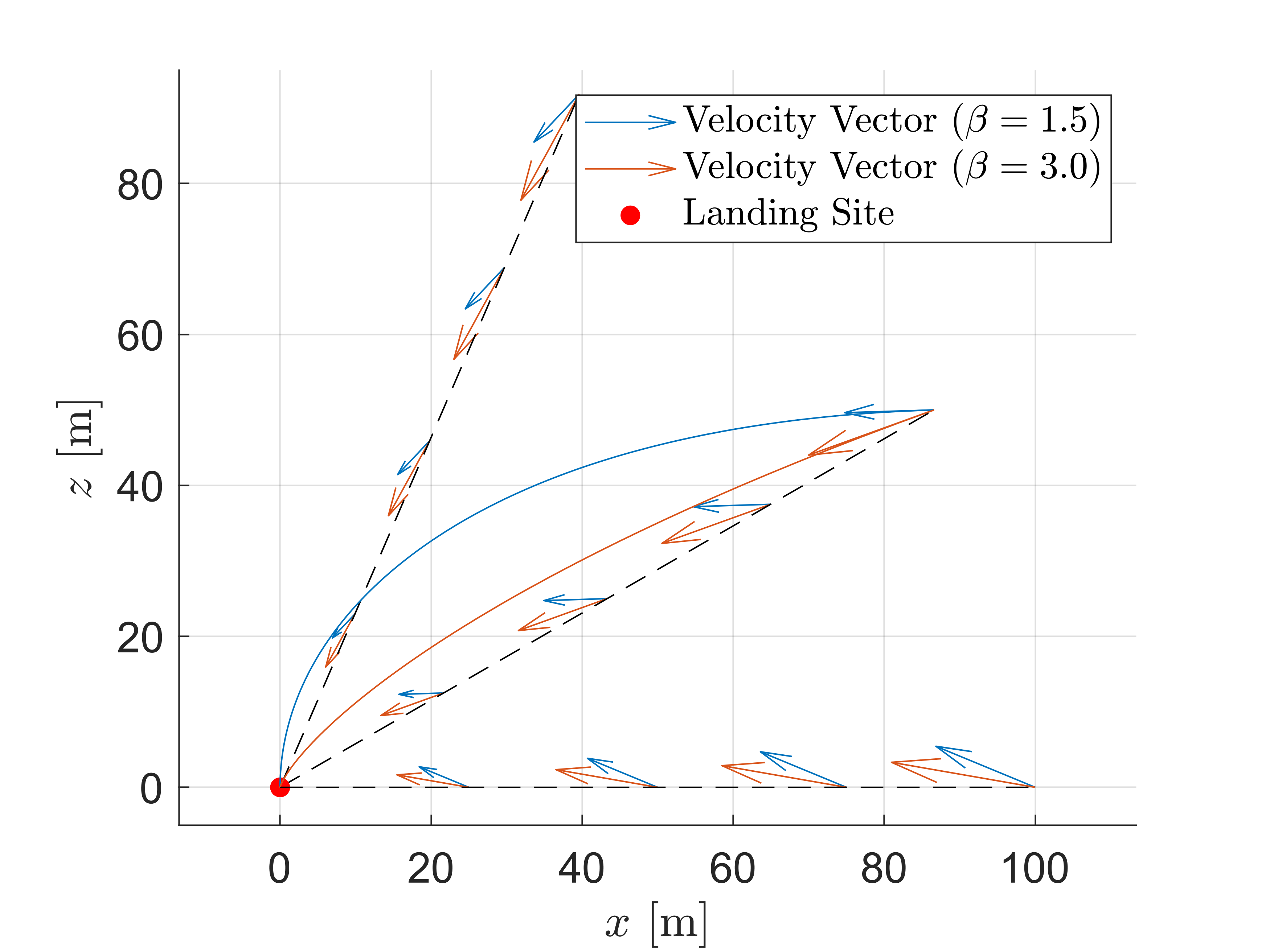}
\caption{(a) Sample velocity vector field and (b) velocity vector along the line with different $\beta$}
\label{fig_2}
\end{figure}

 The trajectory of the vehicle, following the vector field, exhibits a concave-down shape with respect to the landing position. This implies that the vehicle consistently remains above the line of sight (LOS) vector, as depicted in Fig.~\ref{fig_2}. This property is advantageous for the lander as it aids in avoiding potential obstacles around the landing site and ensures the visibility of the camera's field of view \cite{CUI_zemzev, CUI2022313}. This is because, comparing $z_\text{go}/x_\text{go}$ and $s_{\gamma^\ast}/c_{\gamma^\ast}$ gives:
\begin{equation} \label{fpa_los_comp}
\begin{aligned}
\frac{s_{\gamma^\ast}}{c_{\gamma^\ast}} - \frac{z_\text{go}}{x_\text{go}} 
&= \frac{s_{\gamma^\ast}\left(4\beta^2-4\right)\left(2\beta c_{\gamma^\ast} -s_{\gamma^\ast} c_{\gamma^\ast} \right)     -c_{\gamma^\ast}\left(4\beta^2-1\right)\left(2\beta s_{\gamma^\ast} -s^2_{\gamma^\ast} - 1 \right)}
         {c_{\gamma^\ast}\left(4\beta^2-4\right)\left(2\beta c_{\gamma^\ast}-s_{\gamma^\ast} c_{\gamma^\ast} \right)} \\
&= \frac{4\beta^2 + 3(\beta-s_{\gamma^\ast})^2}
         {\left(4\beta^2-4\right)\left(2\beta c_{\gamma^\ast} -s_{\gamma^\ast} c_{\gamma^\ast} \right)} > 0  \\
\end{aligned}
\end{equation}
\noindent proving that $\gamma^\ast$ is always larger than LOS angle as long as the vehicle is on the trajectory.
In addition, larger $\beta$ makes $v^\ast$ larger while $\gamma^\ast$ smaller and vice versa for fixed $x_\text{go}$ and $z_\text{go}$. In other words, the velocity vector field of small $\beta$ makes a steep landing trajectory as shown in Fig.~\ref{fig_2}. Note that the selection of $\beta$ affects not only the shape of the trajectory but also fuel consumption along the trajectory, and the design criteria of $\beta$ will be discussed in the following section. 

Lastly, if the lander precisely follows the vector field of acceleration $\beta$, then the remaining time until landing is:
\begin{equation} \label{t_go_eq}
t_\text{go}(\beta, x_\text{go}, z_\text{go}) = \frac{v^\ast}{(\beta^2-1)g}(\beta - s_{\gamma^\ast})
\end{equation}
\noindent
and the result is used for time-to-go estimation of the guidance law which will be explained in the subsequent section. 

\section{Terminal Landing Guidance Law} \label{sec4}
In this section, we will introduce the guidance law, and the overall guidance logic works as follows: 1) generate a reference trajectory that will guide the lander toward the landing site using the gravity-turn solution, and 2) compute the tracking acceleration that will keep the lander on the reference trajectory.
\subsection{Equation of motion for the lander} \label{sec4.1}
\begin{figure}[b]
	\centering\includegraphics[width=.48\textwidth]{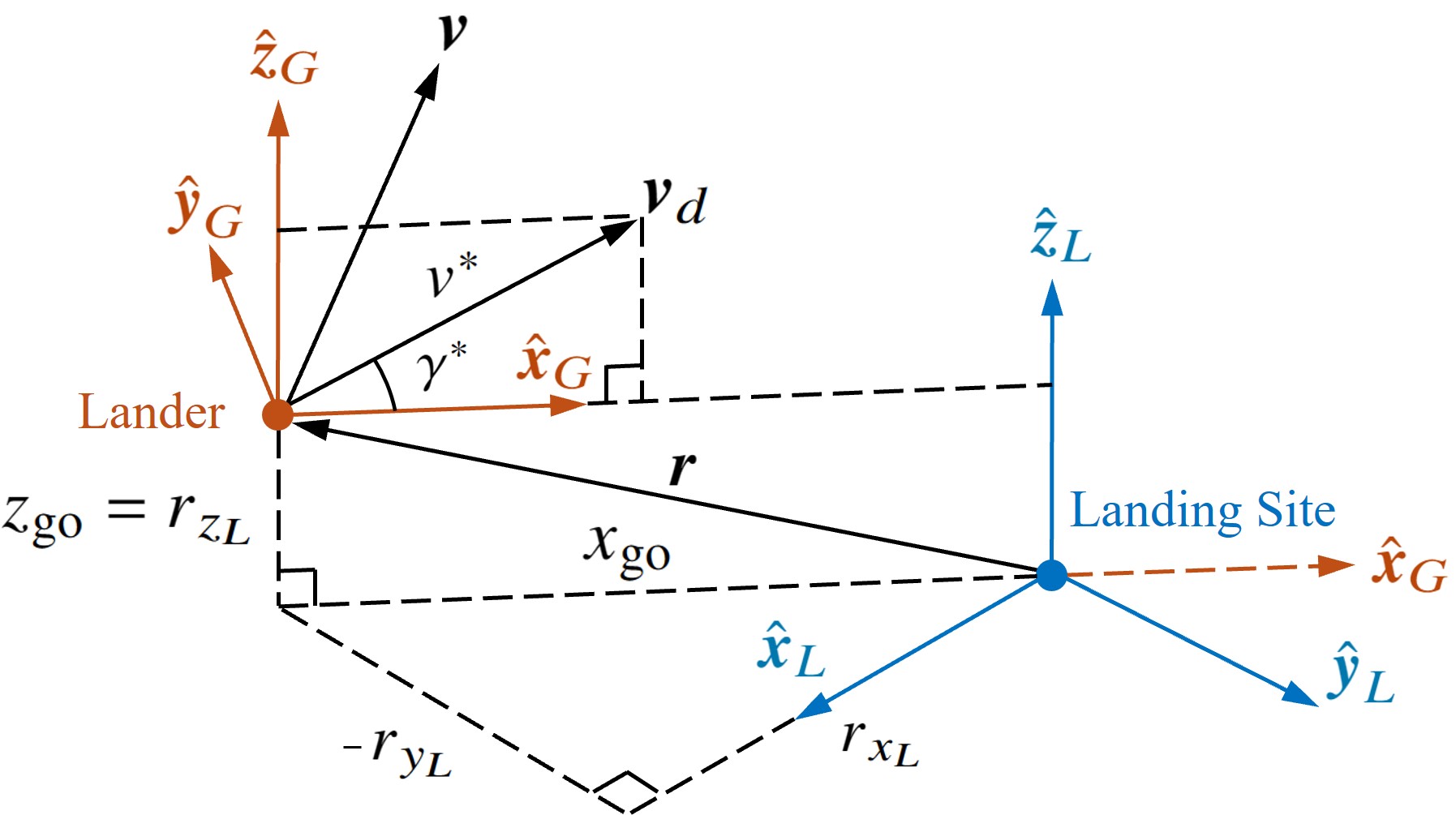}
	\caption{Engagement geometry and guidance frame}
	\label{fig_1}
\end{figure}

Referring to Fig.~\ref{fig_1}, the local reference frame $L$ with basis vectors $\left\{ \hat{\textbf{\textit{x}}}_L, \hat{\textbf{\textit{y}}}_L, \hat{\textbf{\textit{z}}}_L \right\}$ is fixed with respect to the ground surface, where $\hat{\textbf{\textit{z}}}_L$ points towards the zenith, and the landing site is set to be the origin of the frame without loss of generality. In the terminal phase, the Coriolis force can be ignored, and the flat planet model can be reasonably assumed, allowing the gravitational acceleration to be considered constant. Consequently, the motion of the lander can be described by the following three-dimensional point-mass equations.
\begin{subequations}\label{eom_general}
\begin{equation} 
\dot{\textbf{\textit{v}}} = \frac{\textbf{\textit{T}}}{m} + \textbf{\textit{g}} + \textbf{\textit{d}}
\end{equation}
\begin{equation} 
\dot{\textbf{\textit{r}}} = \textbf{\textit{v}}
\end{equation}
\begin{equation}
\dot{m} = - \frac{T}{c}
\end{equation}
\end{subequations}
\noindent where $\textbf{\textit{v}}$ is the velocity vector of the lander, $\textbf{\textit{r}}$ is the position vector of the lander, $\textbf{\textit{T}}$ is the net thrust vector, $\textbf{\textit{g}}$ is the local gravitational acceleration vector, $\textbf{\textit{d}}$ is the total disturbance vector, $m$ is the lander mass, and $c\equiv g_0 I_\text{sp}$ is a effective exhaustive velocity.
The vehicle of interest is assumed to have a thrust-limited engine so the thrust magnitude is constrained as $T_\text{min} < T < T_\text{max}$.

\subsection{Velocity Tracking Control Law} \label{sec4.2}
Referring to Fig.~\ref{fig_1}, we introduce the guidance reference frame $G$, which has basis vectors $\left\{ \hat{\textbf{\textit{x}}}_G, \hat{\textbf{\textit{y}}}_G, \hat{\textbf{\textit{z}}}_G \right\}$. he basis vectors of frame $G$, expressed in frame $L$, are computed as follows:
\begin{equation} \label{G_basis}
\hat{\textbf{\textit{x}}}_G^L = \frac{1}{\sqrt{r_{x_L}^2 + r_{y_L}^2}}\left[-r_{x_L},\  -r_{y_L},\ 0 \right]^\top, \quad \hat{\textbf{\textit{z}}}_G^L = \left[0,\  0,\ 1 \right]^\top, \quad \hat{\textbf{\textit{y}}}_G^L = \hat{\textbf{\textit{z}}}_G^L \times \hat{\textbf{\textit{x}}}_G^L
\end{equation} 
where $r_{x_L}$, $r_{y_L}$, and $r_{z_L}$ are vector elements of $\textbf{\textit{r}}^L= \left[r_{x_L} \ r_{y_L} \ r_{z_L} \right]^\top$ which is the vector $\textbf{\textit{r}}$ expressed in the frame $L$. Then the frame transformation matrix from the $L$ frame to the $G$ frame becomes:
\begin{equation}  \label{dcm_mat}
T_{G/L} = \left[ \hat{\textbf{\textit{x}}}_G^L\quad  \hat{\textbf{\textit{y}}}_G^L\quad \hat{\textbf{\textit{z}}}_G^L
\right]^\top
\end{equation}
From the relative position information, we compute $x_\text{go}$ and $z_\text{go}$ as:
\begin{equation} \label{xgozgo}
    x_\text{go} = \sqrt{r_{x_L}^2 + r_{y_L}^2}, \quad z_\text{go} = -r_{z_L}
\end{equation}
With the current landing geometry determined by $x_\text{go}$ and $z_\text{go}$, we compute the value of $\gamma^\ast$ by solving $h(\gamma) = 0$ in Eq.\eqref{h_eqn} and calculate $v^\ast$ using Eq.\eqref{v_sol}. Then, we can compute the desired velocity vector of the lander, denoted as $\textbf{\textit{v}}_d$, which guides the lander toward the landing site along the reference gravity-turn trajectory.
\begin{equation} \label{vdg}
\textbf{\textit{v}}_d^G 
= \begin{bmatrix} v^\ast \cos{\gamma^\ast} \\ 0 \\ v^\ast \sin{\gamma^\ast} \end{bmatrix}
= \begin{bmatrix} v_x^\ast  \\ 0 \\ v_z^\ast \end{bmatrix}
\end{equation}
where $v_{x}^\ast$ and $v_{z}^\ast$ are the desired velocity components $\textbf{\textit{v}}_d$ expressed in the frame $G$, respectively. 

Let $\textbf{\textit{e}} \equiv \textbf{\textit{v}}_d - \textbf{\textit{v}}$ be the velocity tracking error vector. Then the time derivative of $\textbf{\textit{e}}$ with respect to frame $L$ is:
\begin{equation} \label{error_eqn}
\begin{aligned}
{}^{L}\dot{\textbf{\textit{e}}} &= {}^{L}\dot{\textbf{\textit{v}}}_d - {}^{L}\dot{\textbf{\textit{v}}} \\    
&= {}^{G}\dot{\textbf{\textit{v}}}_d + \boldsymbol{\omega}_{G/L} \times \textbf{\textit{v}}_d - (\textbf{\textit{u}} + \textbf{\textit{g}} + \textbf{\textit{d}} ) \\
\end{aligned}
\end{equation}
where $\boldsymbol{\omega}_{G/L}$ represents the angular velocity vector of the frame $G$ with respect to the frame $L$, and the left superscript on the vector derivative indicates the frame where the derivative is computed. By the chain rule of differentiation, the term ${}^{G}\dot{\textbf{\textit{v}}}_d$ expressed in frame $G$ using Eqs.~\eqref{eqn_13} is equal to:
\begin{equation} 
{}^{G}\dot{\textbf{\textit{v}}}_d^G =  F_{v_d}^{\dagger} F_{r_\text{go}} \textbf{\textit{v}}^G - F_{v_d}^{\dagger} F_{\beta} \dot{\beta} 
\end{equation}
where $F_{v_d}^{\dagger}$ is defined as:
\begin{subequations} \label{Fvd}
\begin{equation} 
F_{v_d}^{\dagger} = 
\frac{1}{\frac{\partial f_x}{\partial v_x^\ast}\frac{\partial f_z}{\partial v_z^\ast} - 
\frac{\partial f_z}{\partial v_x^\ast} \frac{\partial f_x}{\partial v_z^\ast}}
\begin{bmatrix} 
    \frac{\partial f_z}{\partial v_z^\ast}
    & 0 & -\frac{\partial f_x}{\partial v_z^\ast} \\ 0 & 0 & 0 \\ -\frac{\partial f_z}{\partial v_x^\ast} & 0 & \frac{\partial f_x}{\partial v_x^\ast}
\end{bmatrix}
\end{equation}
\begin{equation} 
\frac{\partial f_x}{\partial v_x^\ast} = \frac{1}{v^\ast}\left( 2\beta v_x^\ast {}^2 + 2\beta v^\ast {}^2 - v^\ast v_z^\ast \right)
\end{equation}
\begin{equation} 
\frac{\partial f_x}{\partial v_z^\ast} = \frac{1}{v^\ast}\left( 2\beta v_x^\ast v_z^\ast - v^\ast v_x^\ast \right)
\end{equation}
\begin{equation} 
\frac{\partial f_z}{\partial v_x^\ast} = \frac{1}{v^\ast}\left( 2\beta v_x^\ast v_z^\ast - 2 v^\ast v_x^\ast \right)
\end{equation}
\begin{equation} 
\frac{\partial f_z}{\partial v_z^\ast} = \frac{1}{v^\ast}\left( 2\beta v_z^\ast {}^2 + 2\beta v^\ast {}^2 - 4 v^\ast v_z^\ast \right)
\end{equation}
\end{subequations}
where $F_{v_d}^{\dagger}$ is extended 2 by 2 inverse matrix. $F_{r_\text{go}}$ is defined as:
\begin{subequations} \label{Frgo}
\begin{equation} 
F_{r_\text{go}} = 
\begin{bmatrix} 
    \frac{\partial f_x}{\partial x_\text{go}} & 0 & \frac{\partial f_x}{\partial z_\text{go}} \\ 
    0 & 0 & 0 \\ 
    \frac{\partial f_z}{\partial x_\text{go}} & 0 & \frac{\partial f_z}{\partial z_\text{go}}
    \end{bmatrix}
\end{equation}
\begin{equation} 
\frac{\partial f_x}{\partial x_\text{go}} = -(4\beta^2 -1)g, \quad \frac{\partial f_x}{\partial z_\text{go}} = 0
\end{equation}
\begin{equation} 
\frac{\partial f_z}{\partial z_\text{go}} = -(4\beta^2 -4)g, \quad \frac{\partial f_z}{\partial x_\text{go}} = 0
\end{equation}
\end{subequations} and $F_{\beta}$ is defined as:
\begin{subequations} \label{Fbeta}
\begin{equation} 
F_{\beta} = 
\begin{bmatrix} 
\frac{\partial f_x}{\partial \beta} \\ 0 \\ \frac{\partial f_z}{\partial \beta} 
\end{bmatrix}
\end{equation}
\begin{equation} 
\frac{\partial f_x}{\partial \beta} = 2v^\ast v_x^\ast - 8\beta g x_\text{go}, \quad \frac{\partial f_z}{\partial \beta} = 2v^\ast v_z^\ast - 8\beta g z_\text{go}
\end{equation}
\end{subequations}
Note that $F_\beta$ is required only if a time-varying $\beta(t)$ profile is used and is not necessary for a constant $\beta$ profile. The design criteria for $\beta$ will be discussed in Sec.~\ref{sec4.4}.
For the case of $\boldsymbol{\omega}_{G/L}$, the $\hat{\textbf{\textit{y}}}_G$ velocity component of the lander will make the relative angular velocity as follows \cite{Han_2022}:
\begin{equation} 
\boldsymbol{\omega}_{G/L}^L = \boldsymbol{\omega}_{G/L}^G = \frac{ -(\textbf{\textit{r}}\cdot \hat{x}_G) \hat{\textbf{\textit{x}}}_G^G \times (\textbf{\textit{v}}^G - (\textbf{\textit{v}}\cdot \hat{\textbf{\textit{z}}}_G) \hat{\textbf{\textit{z}}}_G^G)}{(\textbf{\textit{r}}\cdot \hat{\textbf{\textit{x}}}_G)^2}= \begin{bmatrix} 0 \\ 0 \\ -\frac{v_{y_G}}{x_\text{go}} \end{bmatrix}
\end{equation}
where $v_{x_G}$, $v_{y_G}$, and $v_{z_G}$ are vector elements of $\textbf{\textit{v}}^G= \left[v_{x_G} \ v_{y_G} \ v_{z_G} \right]^\top$,
and it can be computed as $ \textbf{\textit{v}}^G= T_{G/L}\textbf{\textit{v}}^L$.

\textbf{\textit{Remark:}} Denominator of $F_{v_d}^{\dagger}$ is always positive, as shown by the following equality:
\begin{equation}
    \frac{\partial f_x}{\partial v_x^\ast}\frac{\partial f_z}{\partial v_z^\ast} - 
\frac{\partial f_z}{\partial v_x^\ast} \frac{\partial f_x}{\partial v_z^\ast} = 6(\beta v^\ast - v_z^\ast)^2 + 2v^\ast {}^2 (\beta^2 - 1) > 0
\end{equation}
\noindent This implies that there exist unique functions $p_x$ and $p_z$ satisfying $p_x(x_\text{go}, z_\text{go}) = v_x^\ast$ and $p_z(x_\text{go}, z_\text{go}) = v_z^\ast$ according to the implicit function theorem, which further validates the existence and uniqueness of solution explained in Sec. \ref{sec3.1}.

With the previously developed results, this paper proposes the velocity tracking acceleration command based on the feedback linearization control law:
\begin{equation} \label{eq_law}
\textbf{\textit{a}}_\text{trk} = {}^{G}\dot{\textbf{\textit{v}}}_d + \boldsymbol{\omega}_{G/L} \times \textbf{\textit{v}}_d - \textbf{\textit{g}} + \frac{k}{\hat{t}_\text{go}}\textbf{\textit{e}} 
\end{equation}

\noindent In the above equation, $k$ is the constant control gain, and the time-to-go estimation $\hat{t}_\text{go}$ is computed as follows:
\begin{equation} \label{tgo_est}
\hat{t}_\text{go} = \frac{1}{(\beta^2-1)g}(\beta v_d - v_z^\ast) + \frac{\lVert\textbf{\textit{e}}\rVert}{ \beta g}
\end{equation}
which has $t_\text{go}$ of Eq. \eqref{t_go_eq} with heuristic correction term accounting for tracking error. If the command is expressed in the frame $G$, one can rearrange the guidance law explicitly as follows:
\begin{equation} \label{eq_law1}
\begin{aligned}
    \textbf{\textit{a}}_\text{trk}^{G} &= F_{v_d}^{\dagger} F_{r_\text{go}} \textbf{\textit{v}}^G - F_{v_d}^{\dagger} F_{\beta} \dot{\beta} + \boldsymbol{\omega}^G_{G/L} \times \textbf{\textit{v}}_d^G - \textbf{\textit{g}}^G + \frac{k}{\hat{t}_\text{go}}\textbf{\textit{e}}^G\\
    &= \left\{ -\beta g \frac{\textbf{\textit{v}}^G_d}{v_d} - F_{v_d}^{\dagger} F_{\beta} \dot{\beta} \right\} + \left\{\frac{k}{\hat{t}_\text{go}}\textbf{\textit{e}}^G \right\} + \left\{-F_{v_d}^{\dagger}F_{r_\text{go}}\textbf{\textit{e}}^G + \Omega \textbf{\textit{e}}^G  \right\} 
\end{aligned}
\end{equation}
\noindent where $\Omega \equiv \text{diag}(0, v_x^\ast/x_\text{go},0)$. It turns out that the tracking command consists of three terms distinguished by braces: the time-varying gravity-turn acceleration, the tracking error feedback acceleration, and the feedforward acceleration. The last expression can be obtained by the relationship $\textbf{\textit{v}} = \textbf{\textit{v}}_d - \textbf{\textit{e}}$ and the fact that $F_{v_d}^{\dagger} F_{r_\text{go}}\textbf{\textit{v}}_d$ is an acceleration on the gravity turn trajectory. The detailed procedure is explained in the Appendix.

\subsection{Characteristics of Velocity Tracking Control Law} \label{sec4.3}
For the purpose of analyzing the properties of the tracking law, $\textbf{\textit{u}} = \textbf{\textit{a}}_\text{trk}$ with a fixed $t_f$ are assumed.

\textbf{\textit{Proposition 3:}} If $k \geq 0$, the final time is fixed, and the system is free from disturbances, then the error dynamics will converge within the desired time.

\noindent
\textbf{\textit{Proof: }} By assumption the desired time is determined as fixed value $t_f$, hence time-to-go is $t_\text{go}=t_f - t$. Substituting the law Eq.~\eqref{eq_law} to the error dynamics Eq.~\eqref{error_eqn} reveals the following closed-loop dynamics.
\begin{equation} 
\dot{\textbf{\textit{e}}}+ \frac{k}{t_\text{go}}\textbf{\textit{e}} = 0
\end{equation}

\noindent
The above differential equation is a typical first-order Cauchy-Euler equation, and the closed-loop solution can be determined as follows: 
\begin{equation} \label{error_sol}
\textbf{\textit{e}} (t) =  \left( \frac{t_\text{go}}{t_{f}} \right)^{k} \textbf{\textit{e}}(t_0)
\end{equation}

\noindent
where $\textbf{\textit{e}}(t_0)$ is the initial tracking error. Therefore, the system is stable, and tracking error converges to zero as $t_\text{go} \to 0$ (i.e $t \to t_f$), if $k \geq 0$.\

\textbf{\textit{Proposition 4:}}
If $k > 1$, and the system is free from disturbances, the tracking acceleration command will converge to the time-varying gravity-turn acceleration of $\beta(t_f)g$ as $t_\text{go} \to 0$. 

\noindent
\textbf{\textit{Proof: }} Substituting the closed-loop solution of error vector Eq.~\eqref{error_sol} into the tracking law Eq.~\eqref{eq_law1} results the following axis acceleration command:
\begin{equation} \label{control_hist}
\textbf{\textit{a}}_\text{trk}^G =  -\beta g \frac{\textbf{\textit{v}}^G_d}{v_d} - F_{v_d} F_{\beta} \dot{\beta}
+ k \frac{t_\text{go}^{k-1}}{t_{f}^k} \textbf{\textit{e}}(t_0)
+ \left(-F_{v_d}F_{r_\text{go}} + \Omega \right)  \left( \frac{t_\text{go}}{t_{f}} \right)^{k} \textbf{\textit{e}}(t_0)
\end{equation}

\noindent Since the matrix term $\left(F_{v_d}F_{r_\text{go}} + \Omega \right)$ is bounded for the entire region and $k > 1$, both feedback and feedforward terms will be diminished as $t_\text{go}\to 0$. Furthermore, if the tracking error is sufficiently small, $v_x^\ast \ll v_z^\ast \to 0$ and $x_\text{go} \ll z_\text{go} \to 0 $ as $t_\text{go} \to 0$ since $\gamma \to -\frac{\pi}{2}$ by properties of gravity turn trajectory. Applying Eq.~\eqref{1d_sol} yields following approximation for $\textbf{\textit{e}} \approx 0$ and $t_\text{go} \to 0$:
\begin{equation} 
F_{v_d}F_\beta \approx 
\begin{bmatrix}
    0 \\ 0 \\ \frac{8\beta g z_\text{go} - 2 |v_z^\ast| v_z^\ast}{4\beta |v_z^\ast| - 4 v_z^\ast}
\end{bmatrix}
\approx 
\begin{bmatrix}
    0 \\ 0 \\ \frac{-4 \frac{\beta}{\beta - 1} |v_z^\ast|v_z^\ast - 2 |v_z^\ast| v_z^\ast}{4\beta |v_z^\ast| - 4 v_z^\ast} 
\end{bmatrix}
\to \textbf{{0}}
\end{equation}

\noindent Therefore, the acceleration command will gradually converge to vertical acceleration as $t_\text{go} \to 0$, if $k > 1$, as desired. 
\begin{equation} \label{last_acc}
\lim_{t_\text{go} \to 0}{ \textbf{\textit{a}}_\text{trk}^G } =  
\lim_{t_\text{go} \to 0} \left[ -\beta g \frac{\textbf{\textit{v}}^G_d}{\textit{v}_d} - F_{v_d}F_\beta \dot{\beta} \right] = 
\begin{bmatrix} 0 \\ 0 \\ \beta(t_f) g \end{bmatrix}
\end{equation}

\textbf{\textit{Proposition 5:}}
If $k > 1$ and the magnitude of disturbances is bounded by the constant value $d_\text{max}$, then the error will converge within the desired time.

\noindent
\textbf{\textit{Proof: }} If there exist disturbances, then the error dynamic becomes
\begin{equation} 
\dot{\textbf{\textit{e}}} + \frac{k}{t_\text{go}}\textbf{\textit{e}} = \textbf{\textit{d}}
\end{equation}

\noindent One can rewrite the equation for each axis since they are decoupled.
\begin{equation} \label{error_decouple}
\dot{e}_i + \frac{k}{t_\text{go}}e_i = d_i \quad i=x,y,z
\end{equation}

\noindent The solution of Eq.~\eqref{error_decouple} is
\begin{equation} 
e_i(t) = \left( \frac{t_\text{go}}{t_{f}} \right)^{k} {e}_i(t_0) + t_\text{go}^{k} \int \frac{d_i(\tau)}{t_\text{go}^{k} (\tau)} d\tau
\end{equation}

\noindent Since the disturbance is bounded, the following inequality holds.
\begin{equation} \label{error_decouples}
\begin{aligned}
\lvert e_i(t)\rvert &= \left| \left( \frac{t_\text{go}}{t_{f}} \right)^{k} {e}_i(t_0) + t_\text{go}^{k} \int \frac{d_i(\tau)}{t_\text{go}^{k} (\tau)} d\tau \right| 
\leq \left( \frac{t_\text{go}}{t_{f}} \right)^{k} \lvert {e}_i(t_0) \rvert + t_\text{go}^{k} \int \frac{ \lvert d_i(\tau) \rvert}{t_\text{go}^{k} (\tau)} d\tau \\
&\leq \left( \frac{t_\text{go}}{t_{f}} \right)^{k} \lvert {e}_i(t_0) \rvert + \frac{t_\text{go}}{{k}-1} d_\text{max} \\
\end{aligned}
\end{equation}

\noindent The result shows that $e_i(t) \to 0$ as $t_\text{go} \to 0$ or equivalently $\textbf{\textit{e}} \to 0$ as $t_\text{go} \to 0$. However, it should be noted that the tracking feedback term in Eq. \eqref{eq_law1} will not converge to exactly zero, but instead, it will have some non-zero value due to the presence of disturbances. As a result, the last-moment acceleration command in Eq. \eqref{last_acc} will have some bias terms in addition to the pure gravity-turn acceleration, which accounts for the non-zero tracking errors caused by disturbances.

\subsection{Selection of Control Parameters} \label{sec4.4}
Up to this point, we have established the necessary conditions for the guidance law, which include $\beta(t) > 1$ and $k > 1$. It is well-known that the optimal thrust profile minimizing fuel consumption is bang-off-bang control, where the profile has at most two switches between the min-max control bounds \cite{Lu_2018, Meditch}. Consequently, if the vehicle is already following the gravity turn trajectory with maximum deceleration, it is indeed in the final bang phase of the fuel-optimal trajectory. Therefore, applying near-maximum deceleration would require less fuel usage while giving additional margin to correct initial handover errors from the previous guidance phase and overcome unknown disturbances. In order to generate the velocity vector, we employ a time-varying deceleration $\beta(t)$ derived with a positive constant ratio $C_\beta < 1$:
\begin{equation} \label{beta_cmd}
\beta(t) = C_\beta \frac{T_\text{max}}{m(t)g}
\end{equation}
Additionally, the following equation is used for $\dot{\beta}$, which is required in Eq.~\eqref{eq_law1}: 
\begin{equation} \label{beta_dot_cmd}
\dot{\beta}(t) = -C_\beta \frac{T_\text{max}}{m^2(t)g} \dot m(t) \approx \frac{\beta^2(t) g}{c} 
\end{equation}
with the assumption that the lander is on the gravity turn trajectory (i.e. $e=0$). Note that $\dot m \approx -\frac{1}{c}\beta(t) m(t) g$ is used instead of directly using $\dot m$. This is because the $\dot m$ is usually not measurable, and estimating it may not be stable.
Effects of selecting $C_\beta$ are illustrated in Fig.~\ref{fig_2} and simulation results in sec~\ref{sec5.2}.

\textbf{\textit{Remark:}} Followings are some comments about determining the $C_\beta$ based on the numerical simulation result:
\begin{itemize}
    \item A $C_\beta \approx 1$ tends to reduce fuel usage but offers no control margin to track the reference trajectory.
    \item Generally, having a 5\% margin, i.e., $C_\beta \approx 0.95$, for control proves to be sufficient to track the reference trajectory.
    \item If $d_\text{max}$ is expected to be less than 10\% of $T_\text{max}$, $C_\beta \approx 0.85$ appears to be appropriate.
    \item If the $m(t)$ is unknown during the flight, using $\beta = \frac{T_\text{max}}{m_0g},\ \dot \beta = 0$ will work with a slight performance drop.
\end{itemize}

The choice of the control gain $k$ determines how quickly the vehicle can track the gravity turn trajectory. Higher gain values lead to early error reduction, which can be beneficial if the vehicle is already near the gravity turn trajectory when the terminal phase starts (i.e. $\lVert\textbf{\textit{e}}(t_0)\rVert$ is small). However, a smaller gain is generally preferred if there are large initial velocity errors since the fuel-optimal trajectory is far from the gravity turn trajectory. While adaptive gain techniques are available \cite{Lee_2018}, simulation results in Sec. \ref{sec5} suggest that a well-selected constant gain $k \approx 2.5$ can achieve satisfactory performance.

\subsection{Ground Collision Avoidance Logic} \label{sec4.5}
In practice, it is essential to have the capability to avoid ground obstacles, which are often modeled as glide slope constraints. The gravity turn trajectory automatically satisfies the constraint due to its concave-down shape, eliminating the need to consider the constraint once the tracking error is sufficiently reduced. 
However, in scenarios where the lander has a large initial hand-over error ($\lVert \textbf{\textit{e}}\rVert \geq C_e$) or has not enough control to track the reference trajectory $\left( \lVert \textbf{\textit{a}}_\text{trk} \rVert \geq T_\text{max}/m(t)\right) $, there is a risk of collision with an obstacle before the tracking error is sufficiently reduced.

The paper proposes the following heuristic condition for computing collision avoidance acceleration $\textbf{\textit{a}}_\text{col}$:
\begin{equation} \label{acol_cond}
\textbf{\textit{a}}_\text{col} = 
\begin{cases} 
 \textbf{0}, & \mbox{if } \lVert \textbf{\textit{e}}(t)\rVert < C_e \ \text{and}\ \lVert \textbf{\textit{a}}_\text{trk}(t) \rVert < T_\text{max}/m(t) \\
 \text{Eq.} \eqref{col_acc}, & \mbox{else }    \\
\end{cases}
\end{equation}
where $C_e$ represents the tracking error threshold. One can choose the threshold as a function of $r$ for better performance or predict the future trajectory numerically to avoid the corner cases. However, based on various simulation results, a constant value for $C_e$ has proven to be sufficient. It's important to note that $C_e$ should be greater than the upper bound of navigation error to avoid unnecessary activation.

\begin{figure}[t]
\centering
\includegraphics[width=.38\textwidth]{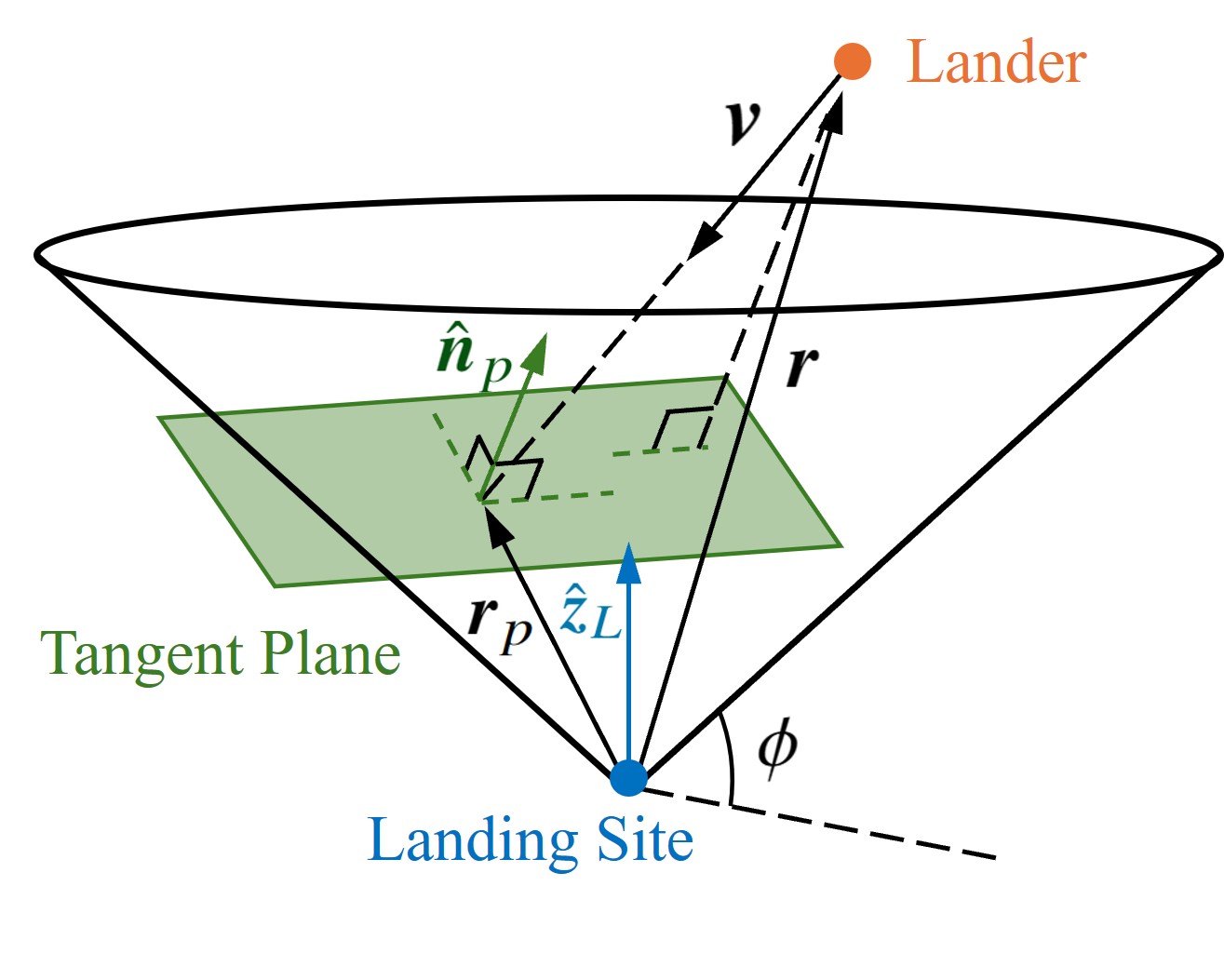} 
\includegraphics[width=.4\textwidth]{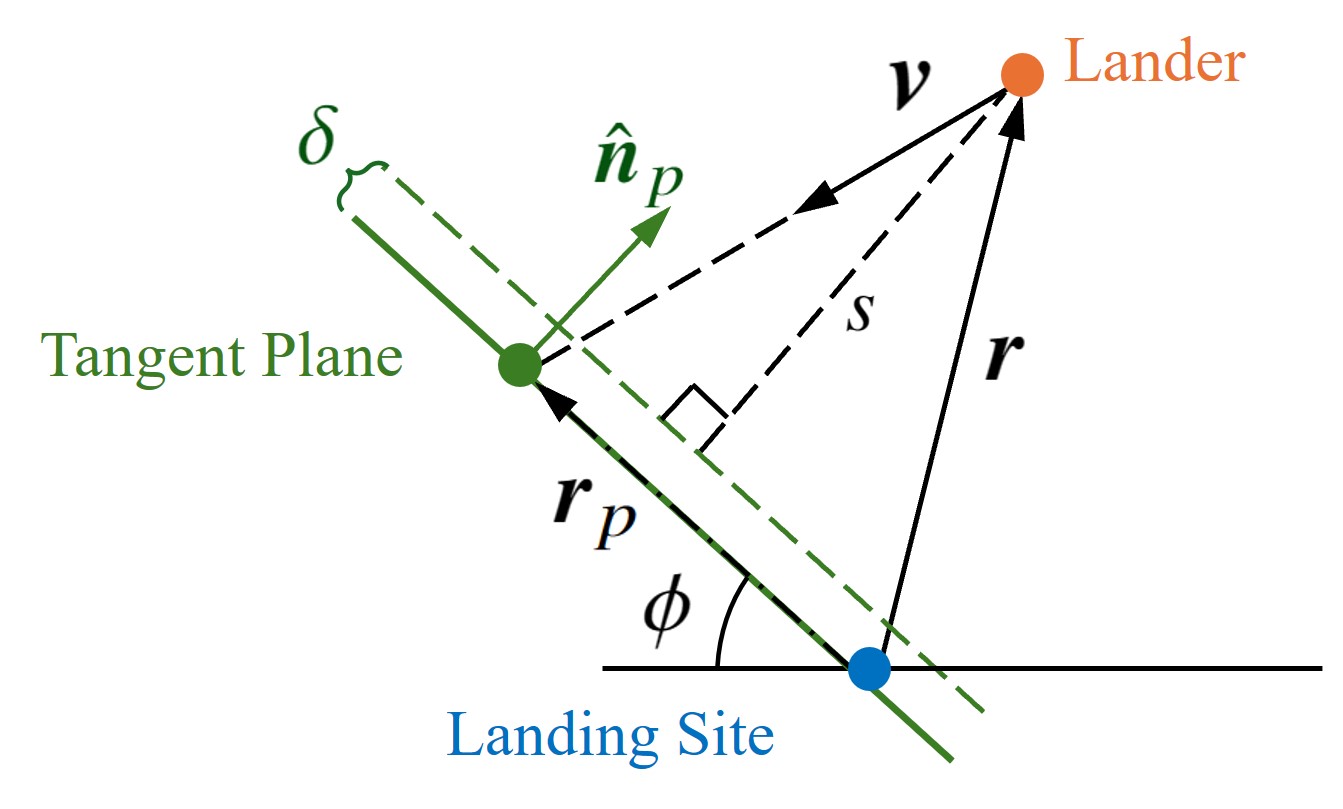}
\caption{Collision Avoidance Geometry (a) A cone constraint (b) Side view of the tangent plane}
\label{fig_obstacle}
\end{figure}
To determine the necessary acceleration for obstacle avoidance, we begin by defining an artificial plane that encapsulates the local surface details of the potential collision point. In this paper, the obstacle created by the glide slope constraint can be modeled as a cone. Thus, using the tangent surface of a cone at the predicted intersection point as an artificial will provide local shape data of a cone. Referring to Fig.~\ref{fig_obstacle}, the predicted intersection point $\textbf{\textit{r}}_p$ can be computed using current position $\textbf{\textit{r}}$ and velocity $\textbf{\textit{v}}$ as:
\begin{subequations} \label{pred_inter_point}
\begin{equation}
\textbf{\textit{r}}_p = \textbf{\textit{r}} + \textbf{\textit{v}} t_p, \quad t_p = \frac{-b_p - \sqrt{|b_p^2 - a_p c_p|}}{a_p}
\end{equation}
\begin{equation} 
a_p = v_{z_L}^2 - (\textbf{\textit{v}}\cdot \textbf{\textit{v}}) \sin^2 \phi, \quad b_p = r_{z_L}v_{z_L} - (\textbf{\textit{r}} \cdot \textbf{\textit{v}})\sin^2 \phi
, \quad c_p = r_{z_L}^2 - (\textbf{\textit{r}}\cdot \textbf{\textit{r}}) \sin^2 \phi
\end{equation}
\end{subequations}
where $\phi > 0 $ is the glide slope angle. Then the normal vector of the tangent surface is computed as:
\begin{equation} \label{plane_normal}
\hat{\textbf{\textit{n}}}_p^L = \frac{1}{\sqrt{ (r_{{p}_x}^2 + r_{{p}_y}^2) \sin^4 \phi +  r_{{p}_z}^2 \cos^4 \phi}  }\begin{bmatrix}
- r_{{p}_x} \sin^2 \phi \\ - r_{{p}_y} \sin^2 \phi  \\ r_{{p}_z} \cos^2 \phi 
\end{bmatrix} 
\end{equation}
where $r_{p_x}$, $r_{p_y}$, and $r_{p_z}$ are vector elements of $\textbf{\textit{r}}_p^L$ in the frame $L$. Note that if $\phi = 0$, i.e. no glide constraint, then the $\hat{\textbf{\textit{n}}}_p^L = \left[0\ 0\ 1\right]^\top$.

Referring to Fig.~\ref{fig_obstacle}, the distance from the lander to the tangent plane with a minimum safety distance $\delta$ is computed as:
\begin{equation} \label{col_acc_d} 
s = \max \left\{ (\textbf{\textit{r}}-\textbf{\textit{r}}_{p})\cdot\hat{\textbf{\textit{n}}}_{p} - \delta,\ \epsilon \right\}
\end{equation}
where $\epsilon\sim0.1$ is a small positive number preventing the division by zero for subsequent computation. Then, the constant deceleration making speed along $\hat{\textbf{\textit{n}}}_p^L$ direction zero, i.e. $\hat{\textbf{\textit{n}}}_p^L \cdot \textbf{\textit{v}} = 0$, after $s$ distance traveling is computed as:
\begin{equation} \label{col_acc_n} 
\textbf{\textit{a}}_{n} =  
\begin{cases} \left( -(\textbf{\textit{g}}\cdot \hat{\textbf{\textit{n}}}_{p}) + \frac{1}{2s}(\textbf{\textit{v}}\cdot\hat{\textbf{\textit{n}}}_{p})^2 \right) \hat{\textbf{\textit{n}}}_{p}, & \mbox{if }                                    \textbf{\textit{v}}\cdot\hat{\textbf{\textit{n}}}_{p} < 0  \\
              \textbf{0}, & \mbox{else }    \\
\end{cases}
\end{equation}
We define a linear sigmoid function $\sigma$ as follows:
\begin{equation} \label{sigmoid} 
\sigma( x, a, b) \equiv 
\begin{cases} 0, & \mbox{if } x < a  \\
              1, & \mbox{if } x > b    \\
              \frac{x - a}{b - a}, & \mbox{else }    \\
\end{cases}
\end{equation}
Then, we compute the collision avoidance acceleration for the tangent plane using Eq.~\eqref{sigmoid} as follows:
\begin{equation} \label{col_acc} 
\textbf{\textit{a}}_{\text{col}}^L = \sigma \left(\lVert\textbf{\textit{a}}_{n}\rVert, \underbar{$C$}_\text{col} \frac{T_\text{max}}{m(t)}, \overline{C}_\text{col} \frac{T_\text{max}}{m(t)} \right) \textbf{\textit{a}}_{n}^L
\end{equation}
where $\underbar{$C$}_\text{col}$ and $\overline{C}_\text{col}$, which are both less than 1, are the lower and upper threshold values for activating the sigmoid function, respectively. Note that $\overline{C}_\text{col} \approx 1$ is preferred to reduce fuel usage for collision avoidance, as known from the fuel-optimal vertical soft landing problem. Compared to the previous work \cite{han2023}, which only considers the $\overline{C}_\text{col}$, adding $\underbar{$C$}_\text{col}$ with the sigmoid function acts as a buffer preventing sudden changes in the acceleration command.

\begin{figure}[b]
    \centering
    \includegraphics[width=\textwidth]{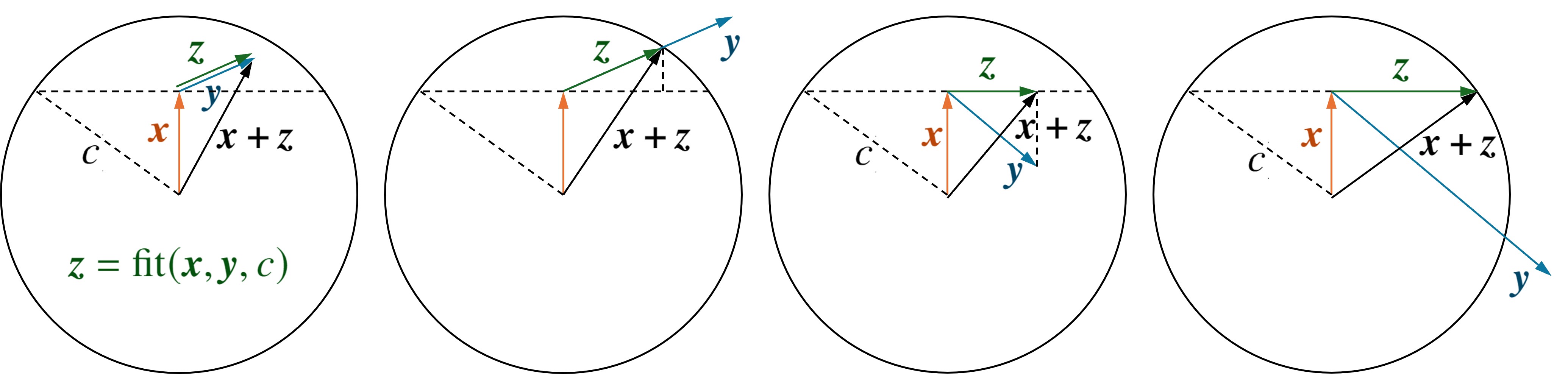}
    \caption{Summation of two vector $\textbf{\textit{x}}+\textbf{\textit{z}}$ using `fit' function, where $\textbf{\textit{z}} = \text{fit}(\textbf{\textit{x}}, \textbf{\textit{y}}, c)$.}
    \label{fig_control_dist}
\end{figure}
\subsection{Thrust Command Saturation Logic} \label{sec4.6}
We first define the two-sided vector magnitude saturation function `sat', which receives a vector and two positive scalars as inputs, as:
\begin{equation} \label{sat_func}
\text{sat}(\textbf{\textit{x}}, a, b) \equiv 
\begin{cases} 
a \hat{\textbf{\textit{x}}}, & \mbox{if } x < a  \\
b \hat{\textbf{\textit{x}}}, & \mbox{if } x > b  \\
\textbf{\textit{x}}, & \mbox{else }\\
\end{cases}
\end{equation}
where $\textbf{\textit{x}}$ is a general vector, $x = \lVert \textbf{\textit{x}} \rVert$, $\hat{\textbf{\textit{x}}} = \textbf{\textit{x}}/x$, $a$ is the lower bound, and $b$ is the upper bound. Referring to Fig.~\ref{fig_control_dist}, we define a function 'fit' that takes two vectors and one positive scalar as inputs, as follows:
\begin{equation} \label{fit_func}
\text{fit}(\textbf{\textit{x}}, \textbf{\textit{y}}, c) \equiv
\begin{cases}
\textbf{0}, & \mbox{if } x > c  \\
\text{sat} \left( \textbf{\textit{y}} - \left(\textbf{\textit{y}} \cdot \hat{\textbf{\textit{x}}} \right) \hat{\textbf{\textit{x}}},\ 0, \sqrt{r^2 - c^2} \right), & \mbox{if }\textbf{\textit{x}} \cdot \textbf{\textit{y}} < 0\ \text{and } x\leq c \\
\text{sat} \left( \textbf{\textit{y}},\ 0,\  \textbf{\textit{x}} \cdot \hat{\textbf{\textit{y}}} + \sqrt{
\left(\textbf{\textit{x}}\cdot \hat{\textbf{\textit{y}}} \right)^2 + c^2 - x^2} \right)
, & \mbox{else }  \\
\end{cases}
\end{equation}
where $\textbf{\textit{x}}$ and $\textbf{\textit{y}}$ are general vectors, $x = \lVert \textbf{\textit{x}} \rVert$ and $c$ is the radius of the sphere. The role of the `fit' function is to modify $\textbf{\textit{y}}$ such that
$\left(\textbf{\textit{x}} + \text{fit}(\textbf{\textit{x}}, \textbf{\textit{y}}, c)\right)\cdot \hat{\textbf{\textit{x}}} \geq \lVert \textbf{\textit{x}} \rVert $
and $\lVert \textbf{\textit{x}} + \text{fit}(\textbf{\textit{x}}, \textbf{\textit{y}}, c)\rVert \leq c$ simultaneously. In simple terms, when we add $\textbf{\textit{x}}$ and $\textbf{\textit{y}}$, we want the resulting vector to preserve the size and direction of $\textbf{\textit{x}}$ by adjusting $\textbf{\textit{y}}$.
$\textbf{\textit{x}} + \textbf{\textit{z}} = \text{fit}(\textbf{\textit{x}}, \textbf{\textit{y}}, c)$

With these two functions, we propose the final acceleration command as follows:
\begin{equation} \label{total_acc} 
\textbf{\textit{u}} = \text{sat} \left(
\textbf{\textit{a}}_\text{col} + \text{fit}\left(\textbf{\textit{a}}_\text{col}, \textbf{\textit{a}}_\text{trk}, \frac{T_\text{max}}{m} \right),\ 
\frac{T_\text{min}}{m},\  \frac{T_\text{max}}{m} \right)
\end{equation}
Firstly, the final 'sat' function is used to ensure the acceleration stays within the minimum and maximum allowable thrust limits. Next, the vector to be saturated is $\textbf{\textit{a}}_\text{col} + \text{fit}\left(\textbf{\textit{a}}_\text{col}, \textbf{\textit{a}}_\text{trk}, T_\text{max}/m \right)$, which is the result of summing $\textbf{\textit{a}}_\text{col}$ and $\textbf{\textit{a}}_\text{trk}$ while preserving size and direction of $\textbf{\textit{a}}_\text{col}$. These two functions allow us to prioritize $\textbf{\textit{a}}_\text{col}$ over $\textbf{\textit{a}}_\text{trk}$ in the vector summation while satisfying to the thrust limits.

\subsection{Summary of Landing Guidance Law}  \label{sec4.7}
The detailed computation sequence of the logic can be outlined as follows:
\begin{enumerate}
    \item Update $\beta(t)$ using Eq.~\eqref{beta_cmd}, $\dot \beta(t)$ using Eq.~\eqref{beta_dot_cmd} and $T_{G/L}$ using Eqs.~\eqref{G_basis} and \eqref{dcm_mat}.
    \item Update $x_\text{go}$ and $z_\text{go}$ using Eq~\eqref{xgozgo}, then compute $\gamma^\ast$ and $v^\ast$ using Eqs.~\eqref{eqn_16} and ~\eqref{v_sol}.
    \item Compute $\textbf{\textit{v}}_d^G$ using Eq.~\eqref{vdg} and $\textbf{\textit{e}}^G = \textbf{\textit{v}}_d^G - \textbf{\textit{v}}^G$ then update $\hat{t}_\text{go}$ using Eq.~\eqref{tgo_est}.
    \item Compute $F_{v_d}^{\dagger}$, $F_{r_\text{go}}$ and $F_\beta$ using Eqs.~\eqref{Fvd} to \eqref{Fbeta}, then 
    $\textbf{\textit{a}}_\text{trk}^{L} = T_{G/L}^\top\textbf{\textit{a}}_\text{trk}^{G}$ using Eq.~\eqref{eq_law1}. 
    \item Compute $\textbf{\textit{a}}_\text{col}^L$ based on the heuristic condition in Eq.~\eqref{acol_cond}. 
        \begin{itemize}
            \item If $\lVert \textbf{\textit{e}}\rVert < C_e$ and $\lVert \textbf{\textit{a}}_\text{trk} \rVert < T_\text{max}/m$, set 
            $\textbf{\textit{a}}_\text{col}^L = \textbf{0}$
            \item Else, compute $\hat{\textbf{\textit{n}}}_p^L$ from Eqs.~\eqref{pred_inter_point}-\eqref{plane_normal}, 
            $\textbf{\textit{a}}_n^L$ using Eqs.~\eqref{col_acc_d}-\eqref{col_acc_n}, then $\textbf{\textit{a}}_\text{col}^L$ as Eq.~\eqref{col_acc}
        \end{itemize}
    \item Compute control command $\textbf{\textit{u}}^L$ with $\textbf{\textit{a}}_\text{trk}^L$ and $\textbf{\textit{a}}_\text{col}^L$ using Eq.~\eqref{total_acc}
\end{enumerate}
and the above sequence of steps needs to be performed during every iteration of the guidance process.

\begin{table}[b!]
	\fontsize{10}{10}\selectfont
    \caption{Lander Parameters Used for Numerical Simulations} \label{tab:common_para} \centering 
    \begin{tabular}{c  r  r  r } 
      \hline 
      Parameters    & Description & Value & Unit\\
      \hline 
      g      & Local gravitational acceleration of Mars & 3.7114 & m/s$^2$ \\
      $m_\text{wet}$   & Mass of lander including fuel & 1905 & kg \\
      $m_\text{dry}$   & Mass of lander excluding fuel & 1405 & kg \\
      $T_\text{max}$   & Upper bound of thrust (80\%) & 13258 &  N\\
      $T_\text{min}$   & Lower bound of thrust (30\%) & 4971.8 & N \\
      $c$              & Effective exhaust velocity $(g_0 I_\text{sp})$  & $1965$ & m/s \\
      \hline
   \end{tabular}
\end{table}
\section{Numerical Simulation} \label{sec5}
In this section, we will conduct numerical simulations to verify the performance and robustness of the proposed guidance law.  The lander parameters introduced in \cite{Acikmese_2007} are used for the simulations, and these parameters are summarized in Table \ref{tab:common_para}. Additionally, the control parameters of the proposed logic are summarized in Table \ref{tab:control_para} as well.  During the simulations, the termination condition is set to be the distance from the lander to the target being less than $0.01$ m and the speed of the lander being less than $0.05$ m/s. Once this condition is met, the simulation is terminated.

\begin{table}[t!]
	\fontsize{10}{10}\selectfont
    \caption{GT Control Parameters Used for Numerical Simulations} \label{tab:control_para} \centering 
    \begin{tabular}{c  r  r  r } 
      \hline 
      Parameters    & Description & Value & Unit\\
      \hline 
      $k$            & Tracking control gain of each axis  & 2.4 & - \\
      $C_\beta$        & Ratio for reference trajectory acceleration  & 0.95 & - \\
      $C_e$            & Velocity tracking error threshold & 20 & m/s \\
      $\delta$         & Safety distance margin for collision avoidance & 5 & m \\
      $\overline{C}_\text{col}$   & Collision avoidance logic trigger threshold  & 0.95 & - \\
      $\underbar{$C$}_\text{col}$   & Collision avoidance logic trigger threshold  & 0.75 & - \\
      \hline
   \end{tabular}
\end{table}

\begin{table}[b]
\caption{Results Summary of Representative Scenarios \label{tab:case_table}}
\centering
\resizebox{\columnwidth}{!}{%
\begin{tabular}{l cccc cccc ccc}
\hline
    & \multicolumn{3}{c}{Case 1} & & \multicolumn{3}{c}{Case 2} & & \multicolumn{3}{c}{Case 3} \\
    \cline{2-4} \cline{6-8} \cline{10-12} 
Method & $\Delta m $ [kg] & $\theta_u$ [deg]  & $\gamma_f$ [deg] && $\Delta m $ [kg] & $\theta_u$ [deg] & $\gamma_f$ [deg] && $\Delta m $ [kg] & $\theta_u$ [deg]  & $\gamma_f$ [deg] \\
\hline
GT          &  246.62 & 88.55 & -89.32 && 390.16 & 87.46 & -88.43 && 410.39 & 88.32 & -88.65\\
ZEM/ZEV     &  254.98 & 53.01 & -22.98 && 421.72 & 47.59 & -12.53 && 417.08 & 37.80 & 2.31\\
TZEM/ZEV    &  254.98 & 53.01 & -15.25 && 421.73 & 47.60 & -5.09 && 415.35 & -22.23 & -36.41\\
FD${}^2$PG  &  257.00 & 88.02 & -86.73 && 421.73 & 88.19 & -88.71 && 418.09 & 87.93 & -88.09\\
OPT         &  237.39 & 44.78 & -18.52 && 380.33 & 35.82 & -11.09 && 398.31 & 33.38 & -9.09 \\
\hline
\end{tabular}
}
\end{table}

\subsection{Representative Scenario Analysis} \label{sec5.1}
Three different scenarios are tested to cover a wide range of initial conditions. Each case represents `Small Initial Offset', `Large Initial Offset', and `Worst Initial Condition', respectively. In addition, three different feedback type control laws will be compared in this study:  Zero-Effort-Miss/Zero-Effort-Velocity Guidance (ZEM/ZEV) \cite{Souza_opt}, Two-Phase Zero-Effort-Miss/Zero-Effort-Velocity Guidance (TZEM/ZEV)\cite{Wang_twophase}, and Fractional-Polynomial Guidance (FP${}^2$DG)\cite{Lu_fractional}.
It is important to note that FP${}^2$DG does not provide explicit logic to compute time-to-go but use a value close to that of the fuel optimal trajectory, which is not a practical approach. In contrast, ZEM/ZEV and TZEM/ZEV provide a method to compute time-to-go of energy optimal trajectory in Eq.\eqref{tgo_opt}, and this time-to-go estimation is also applied to FP${}^2$DG. All control parameters used are written in their paper if explicitly mentioned. For FP${}^2$DG method, $\gamma = 1$, $k_r = 8$, $\textbf{\textit{a}}_{T_f}^\ast = -2\textbf{\textit{g}}$ are selected from various pair, and for TZEM/ZEV the forward Euler integration method with step size 0.1-second is used for collision detection.
\begin{equation} \label{tgo_opt}
\frac{g^2}{2}t_\text{go}^4 - 2\left(\textbf{\textit{v}}\cdot \textbf{\textit{v}} \right) t_\text{go}^2 - 12 \left( \textbf{\textit{v}} \cdot \textbf{\textit{r}} \right) t_\text{go} - 18 \left( \textbf{\textit{r}} \cdot \textbf{\textit{r}} \right) = 0
\end{equation}
Lastly, the results of a fuel optimal trajectory (OPT), obtained by GPOPS, are included for comparison, providing insights into computational guidance methods that closely approximate the optimal trajectory\cite{gpops}. The fuel consumption ($\Delta m$), elevation angle of acceleration command with respect to frame $L$ ($\theta_u \equiv \sin^{-1}(u_z/\lVert\textbf{\textit{u}}\rVert)$), and the flight path angle $\gamma_f$ at the last moment of each method are summarized in Table. \ref{tab:case_table}.

\begin{figure}[!ht]
\centering
\includegraphics[width=.46\textwidth]{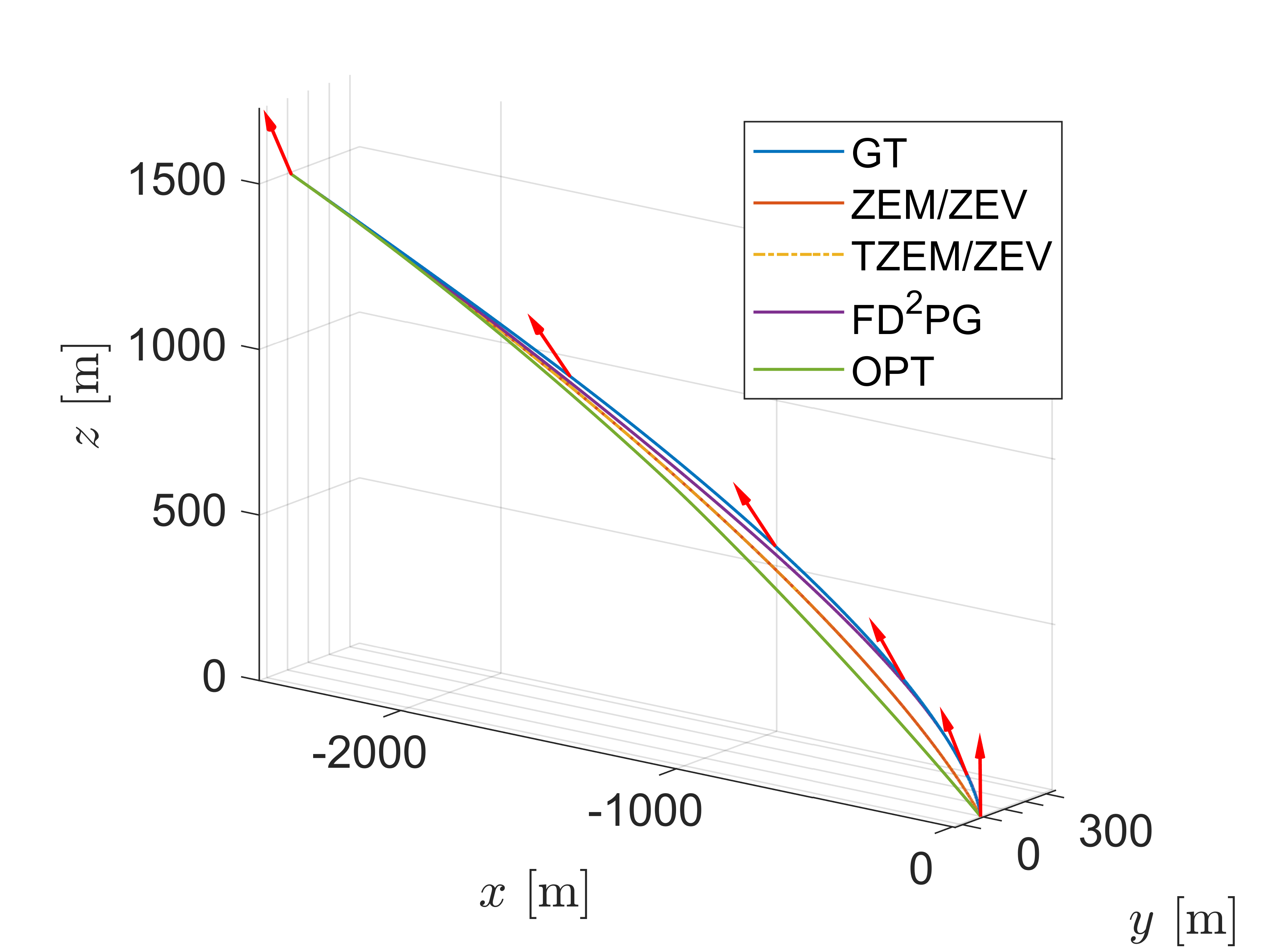} 
\includegraphics[width=.46\textwidth]{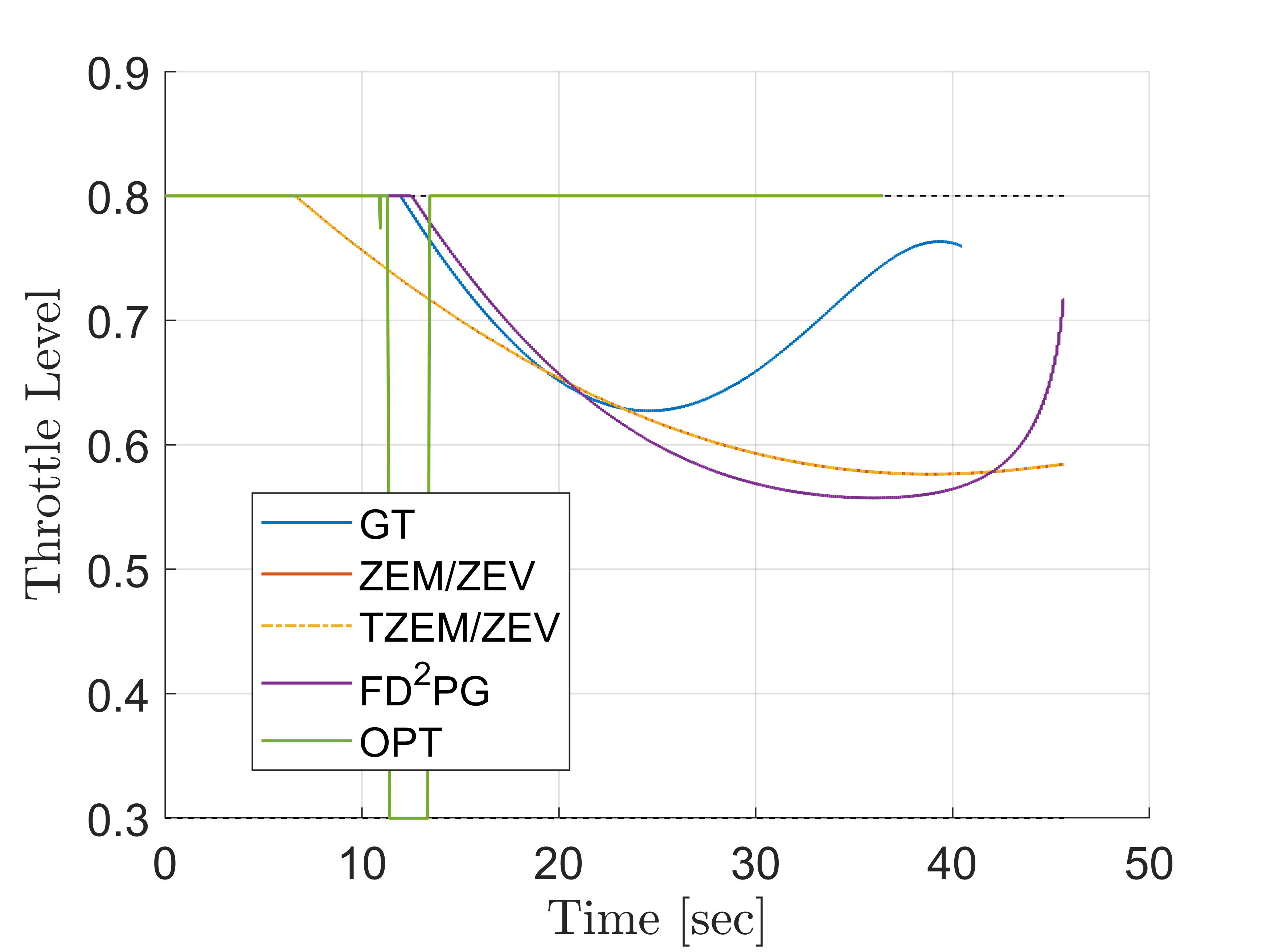}
\includegraphics[width=.46\textwidth]{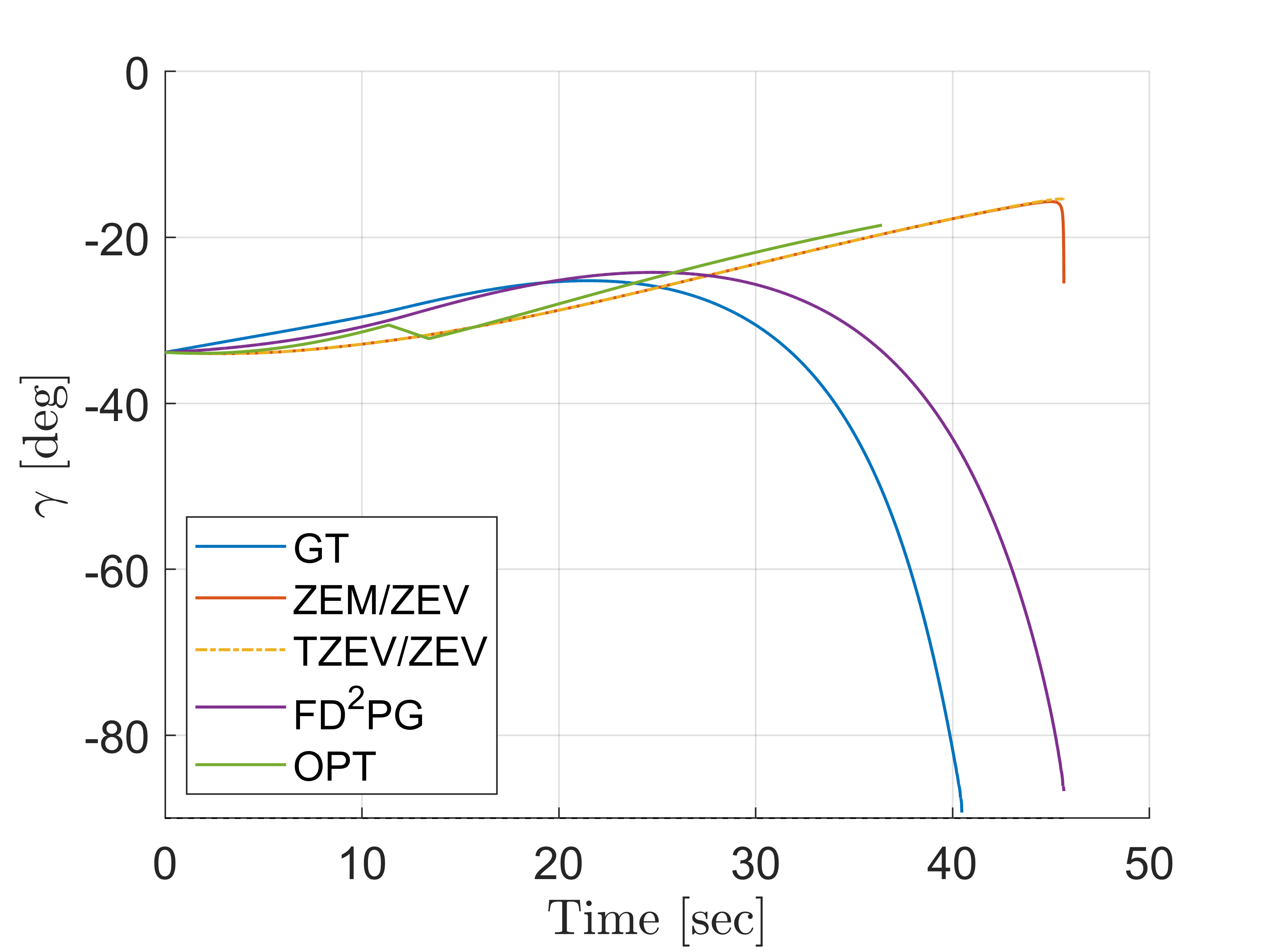} 
\includegraphics[width=.46\textwidth]{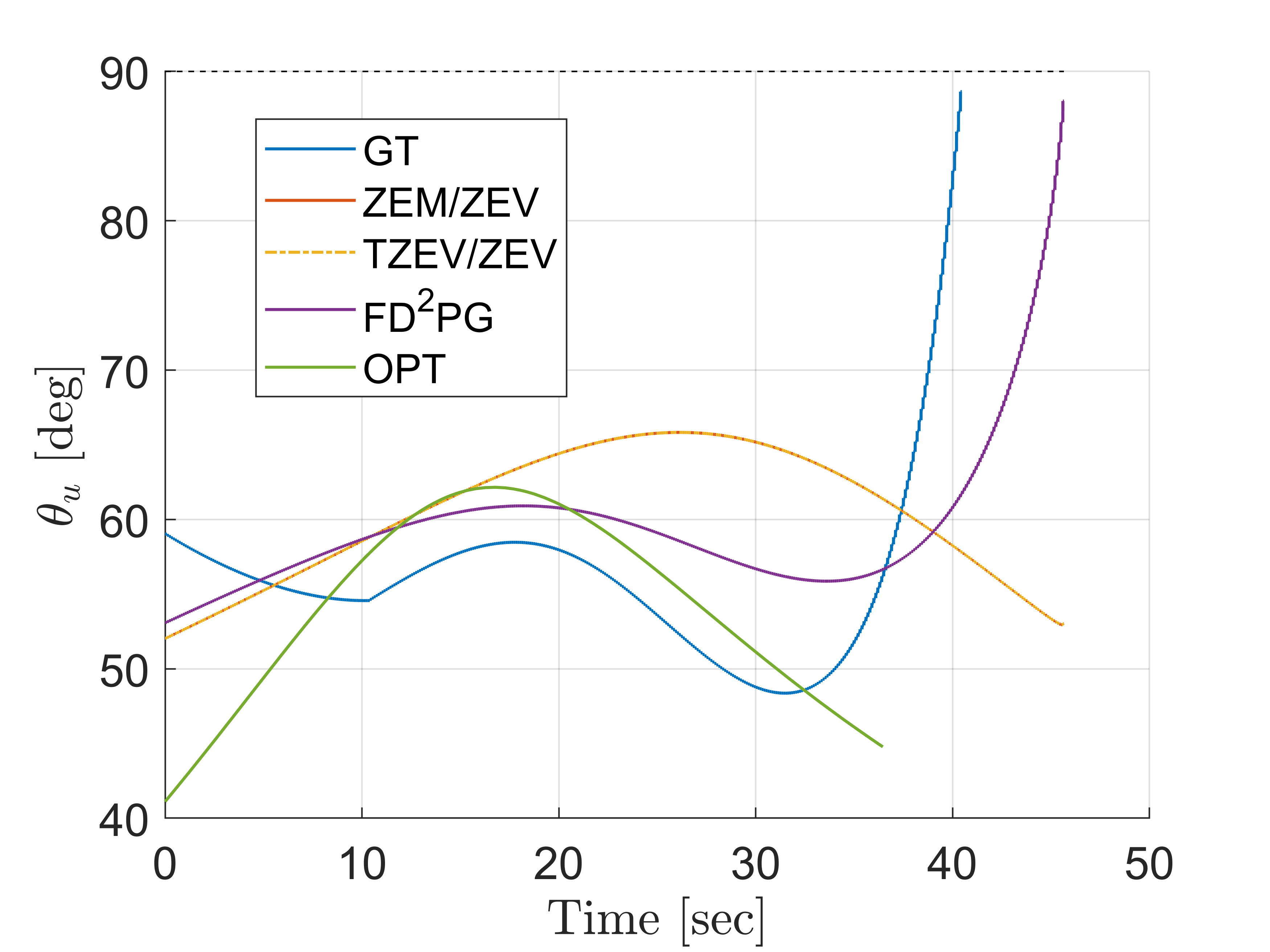}
\caption{Results of Scenario 1 (a) Trajectory (b) Throttle Level (c) Flight-Path Angle (d) Elevation angle $\theta_u$}
\label{fig_case1}
\end{figure}

\textbf{\textit{Scenario 1: Small Initial Deviation}} \\
The initial states for the scenario are $\textbf{\textit{r}}_0^L = \left[-2500,\ 0,\ 1500 \right]^T \text{m}$ and $\textbf{\textit{v}}_0^L = \left[100,\ 50,\ -75 \right]^T \text{m/s}$, which has slight deviation on an initial flight direction. Figure.~\ref{fig_case1}.a describes the lateral view of trajectories of each algorithm, and red arrows indicate the thrust direction of GT logic along the trajectory. It can be observed that all algorithms exhibit similar trajectories including the optimal trajectory, and the GT method requires the least amount of fuel compared to the other feedback algorithms. f the trajectory of ZEM/ZEV is free from collision, TZEM/ZEV is identical to ZEM/ZEV. Additionally, Fig.~\ref{fig_case1}.b presents the throttle level profiles of each algorithm. Although GT law originated from 2D motion, combined control law can handle practical 3D landing cases well. In addition, Figures \ref{fig_case1}.c and \ref{fig_case1}.d show that GT and FP${}^2$DG achieved a vertical landing at the last moment, making the trajectory concave down.

\begin{figure}[!t]
\centering
\includegraphics[width=.46\textwidth]{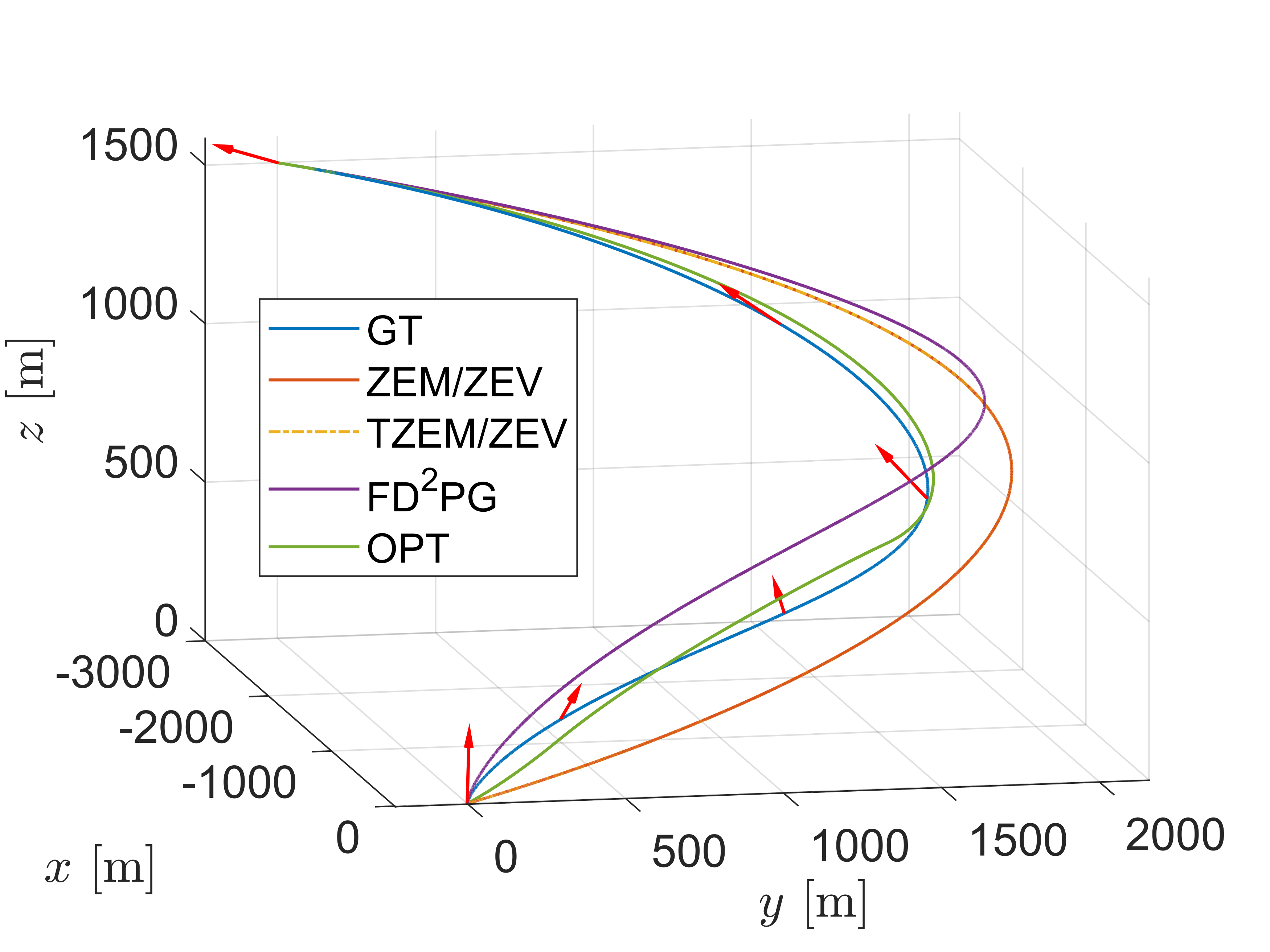} 
\includegraphics[width=.46\textwidth]{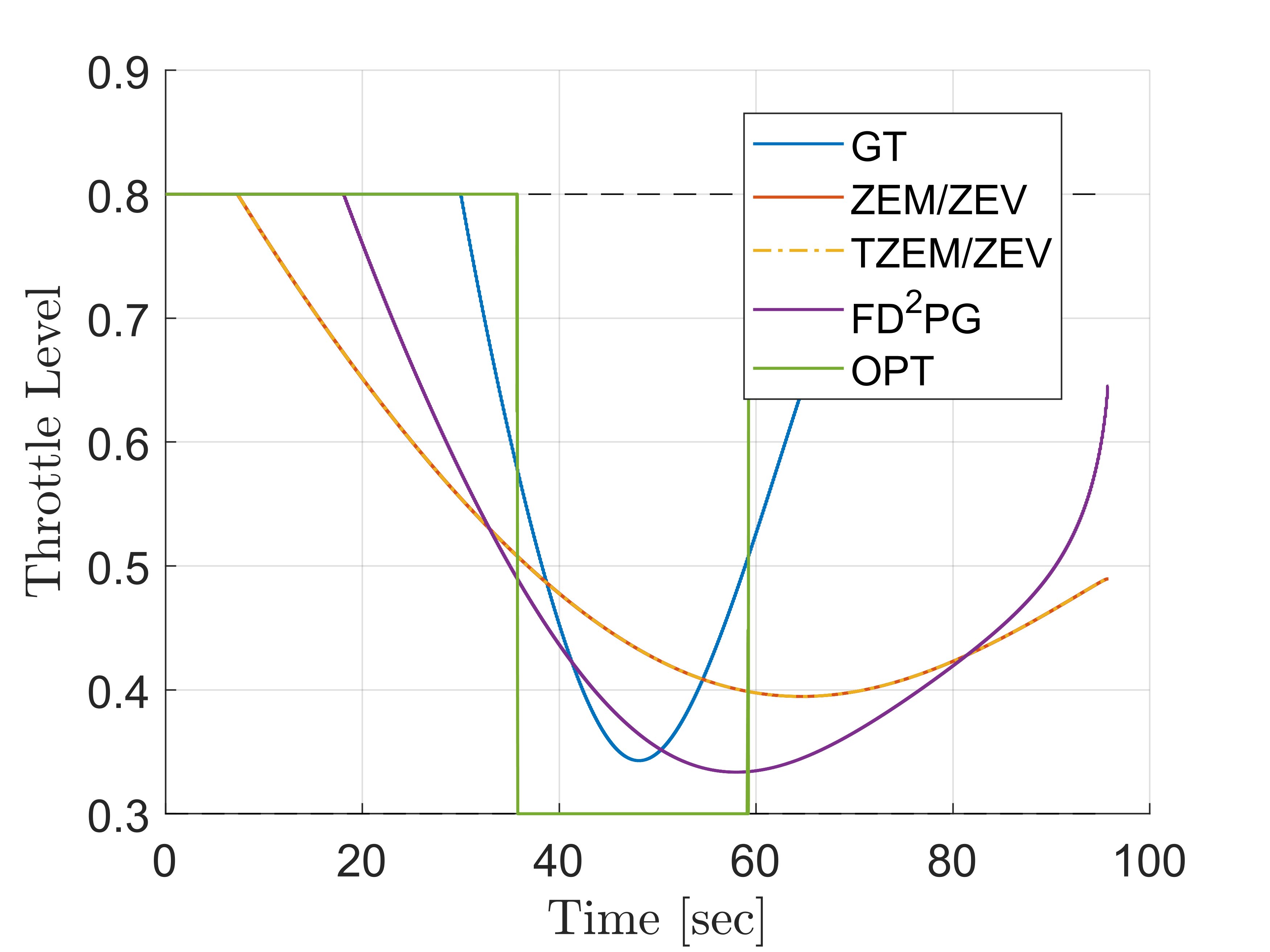}
\includegraphics[width=.46\textwidth]{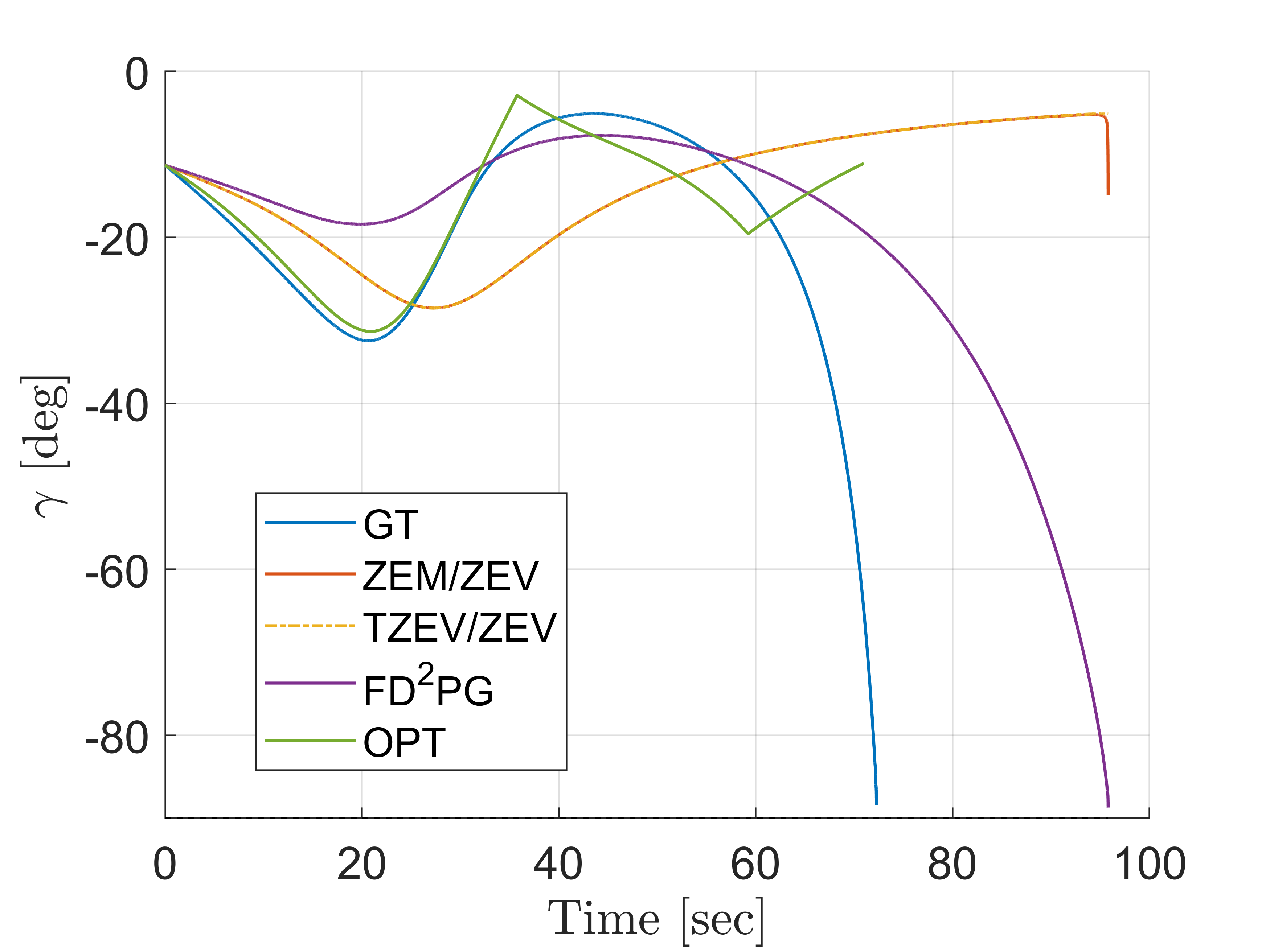} 
\includegraphics[width=.46\textwidth]{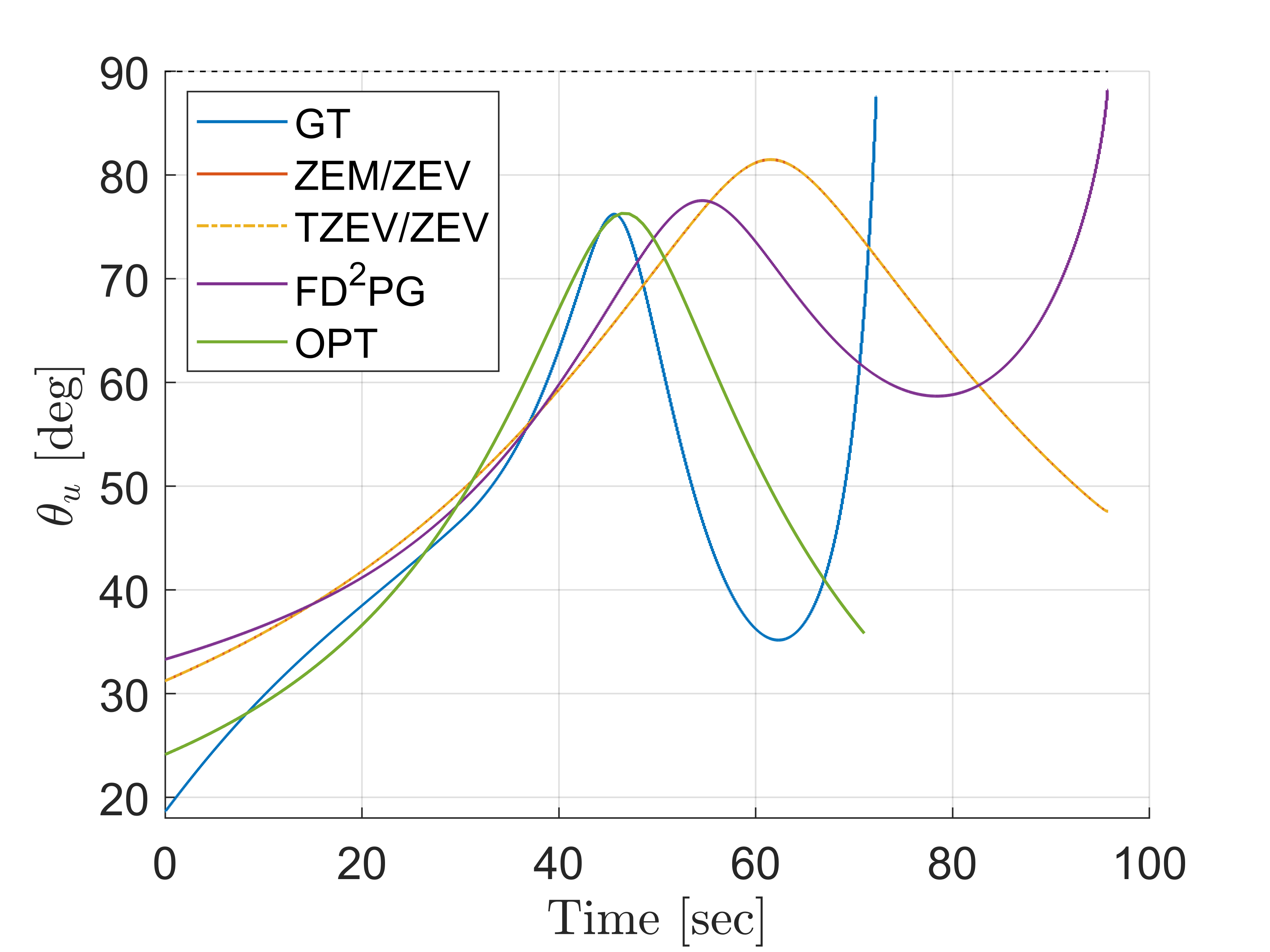}
\caption{Results of Scenario 2 (a) Trajectory (b) Throttle Level (c) Flight-Path Angle (d) Elevation angle $\theta_u$}
\label{fig_case2}
\end{figure}

\textbf{\textit{Scenario 2: Large Initial Deviation}} \\
The initial states for the scenario are $\textbf{\textit{r}}_0^L = \left[-3000,\ 0,\ 1500 \right]^T \text{m}$ and $\textbf{\textit{v}}_0^L = \left[0,\ 150,\ -30 \right]^T \text{m/s}$, which has 90 degrees of heading error with large positional displacement, and Fig. \ref{fig_case2}.a, \ref{fig_case2}.b, \ref{fig_case2}.c, and \ref{fig_case2}.d depict 3D trajectory, throttle level, thrust elevation angle, and flight path angle of each algorithm. Despite the large initial error, the GT method successfully landed at the target site with efficient fuel usage. Similar to Scenario 1, GT and FP${}^2$DG show a vertical thrust direction at the last moment, and TZEM/ZEV is identical to ZEM/ZEV. Note that $\theta_u \approx 90^\circ$ and $\gamma_f \approx -90^\circ$ imply a very small thrust offset angle at the landing moment.

\begin{figure}[!t]
\centering
\includegraphics[width=.46\textwidth]{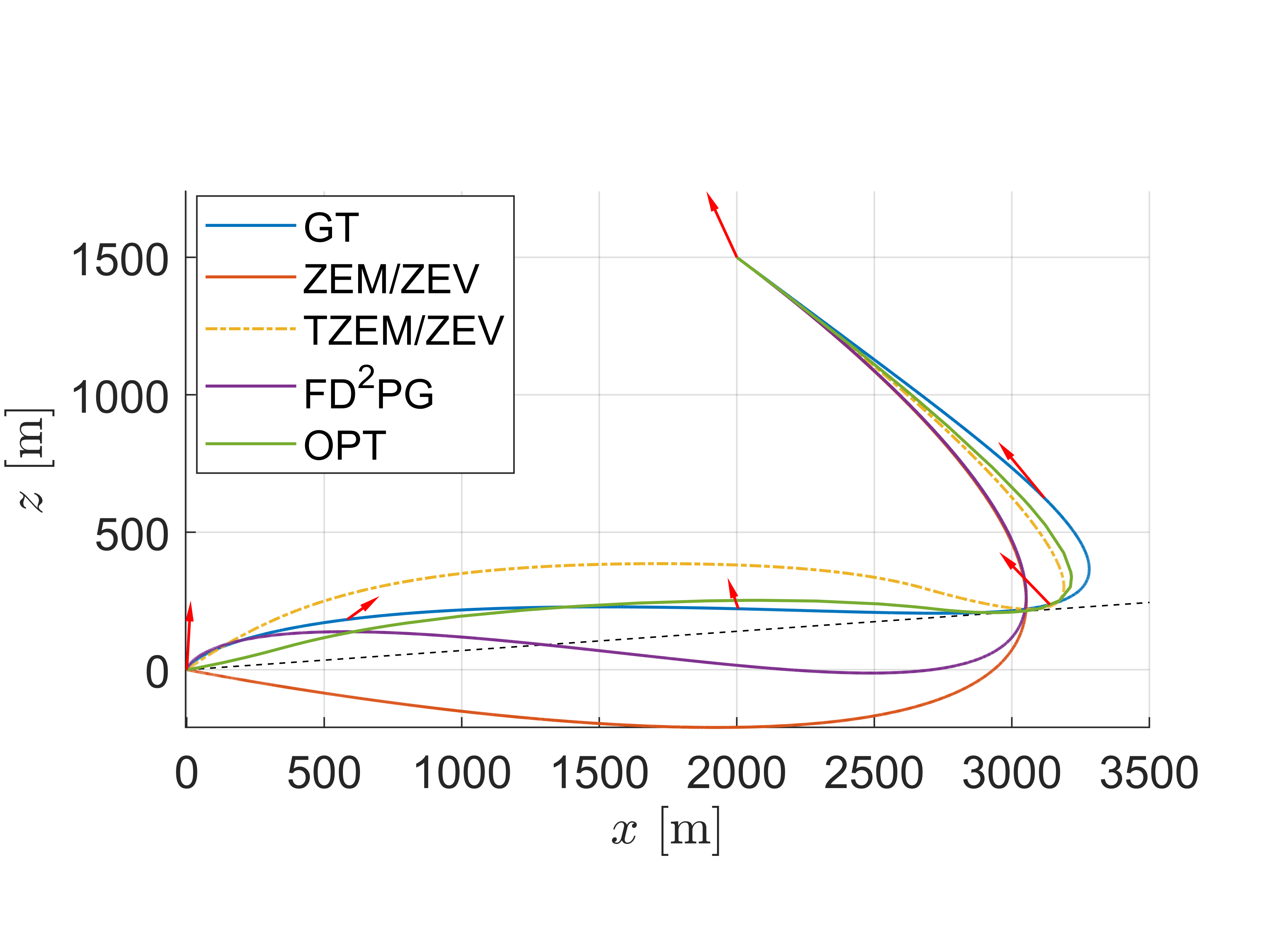} 
\includegraphics[width=.46\textwidth]{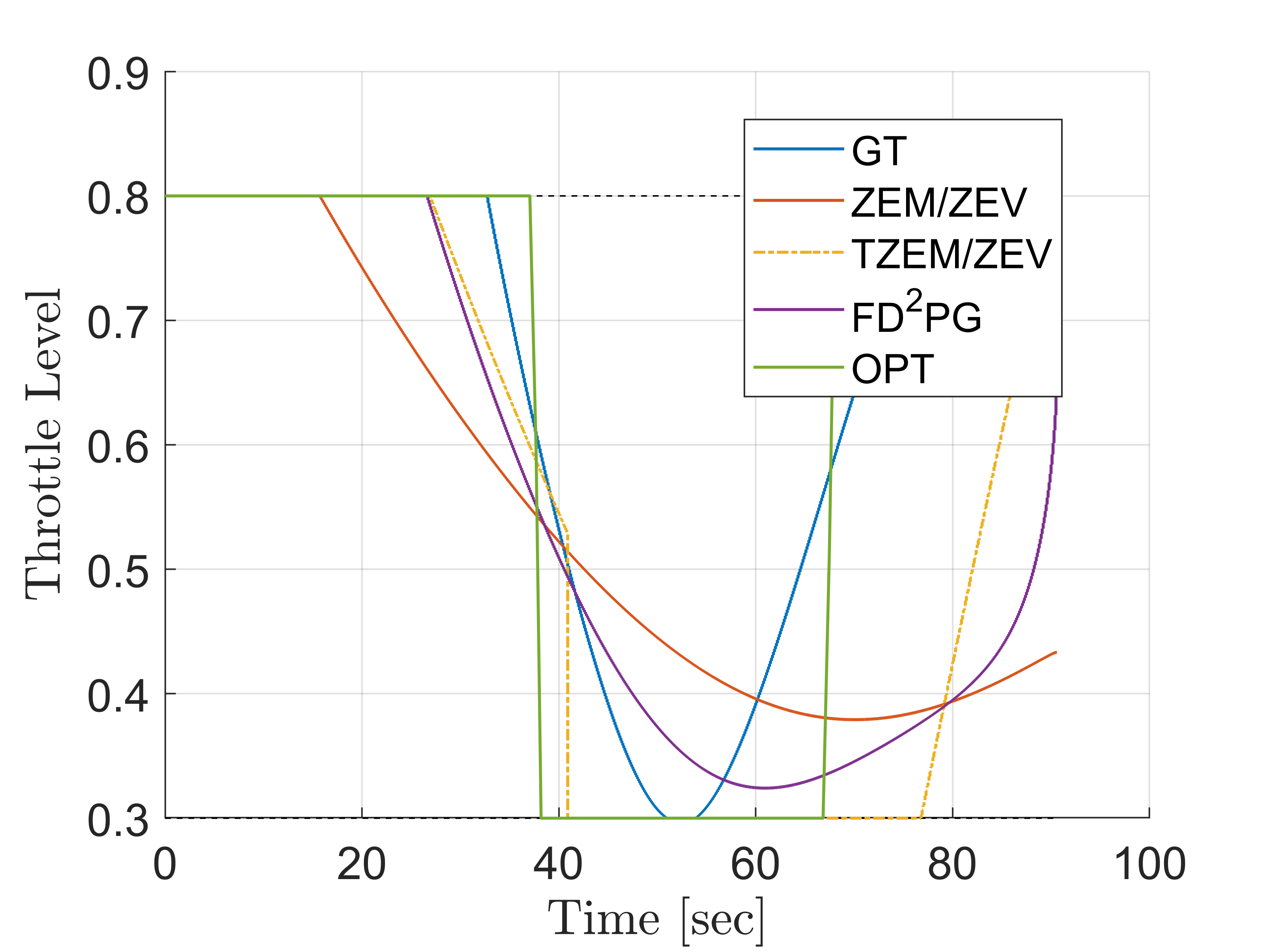}
\includegraphics[width=.46\textwidth]{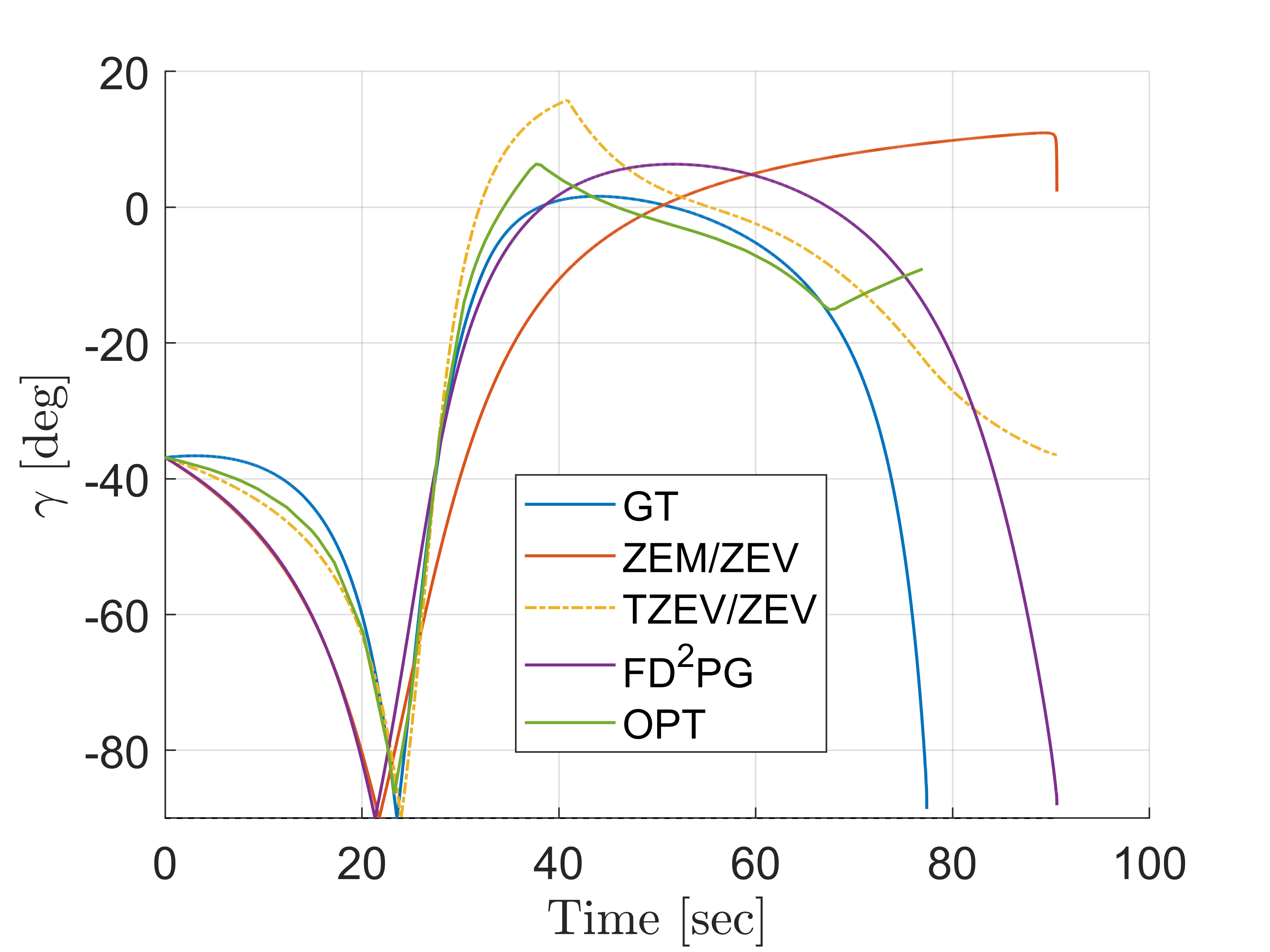} 
\includegraphics[width=.46\textwidth]{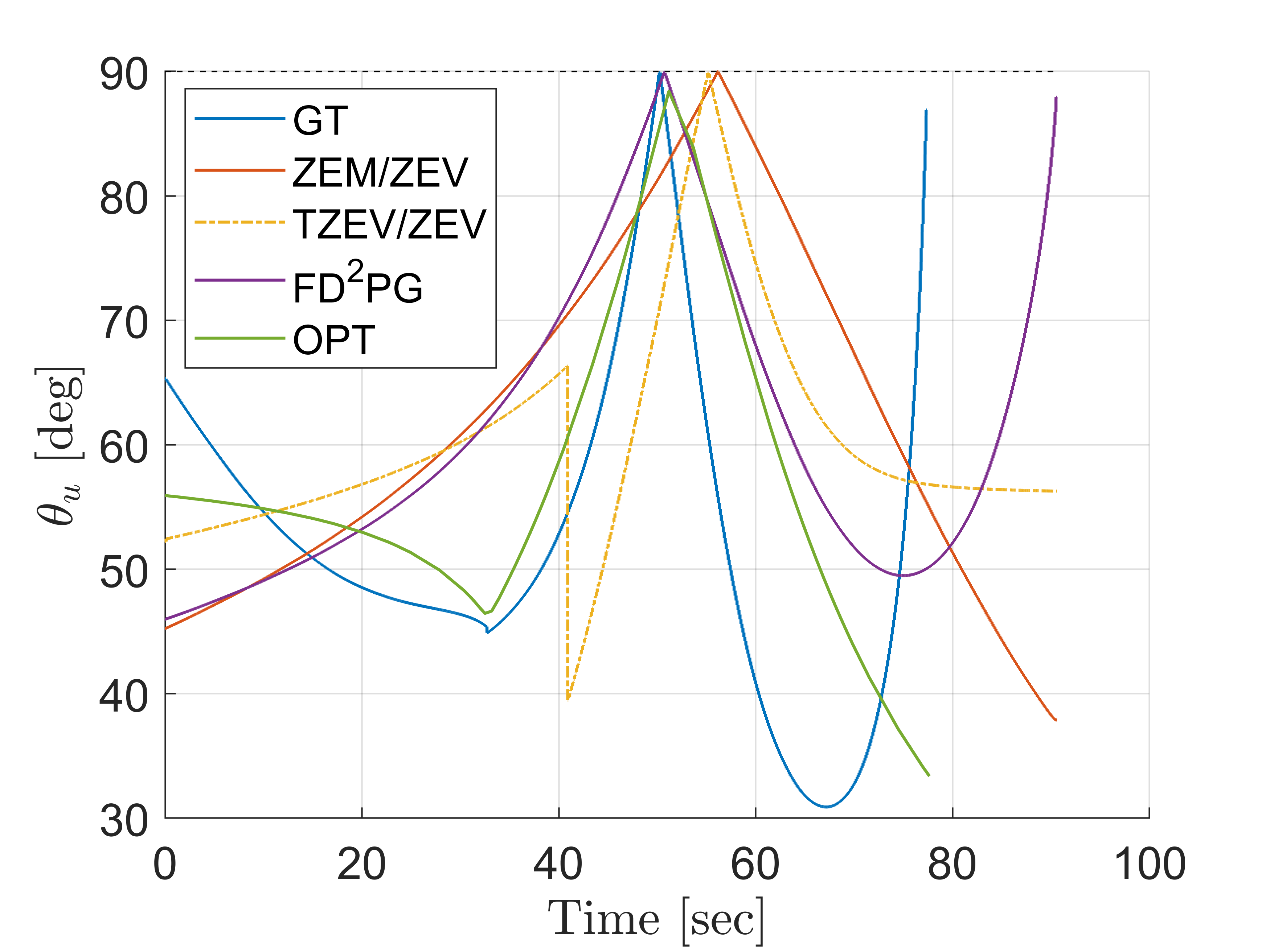}
\caption{Results of Scenario 3 (a) Trajectory (b) Throttle Level (c) Flight-Path Angle (d) Elevation angle $\theta_u$}
\label{fig_case3}
\end{figure}

\textbf{\textit{Scenario 3: Overshoot Initial Speed}} \\
The initial states for the scenario are $\textbf{\textit{r}}_0^L = \left[2000,\ 0,\ 1500 \right]^T \text{m}$ and $\textbf{\textit{v}}_0^L = \left[100,\ 0,\ -75 \right]^T \text{m/s}$. This scenario represents the worst-case benchmark scenario with a 4-degree glide slope constraint, which is represented as a black dashed line in Fig. \ref{fig_case3}. The GT method and TZEM/ZEV were able to satisfy the constraint due to collision avoidance logic. Although FP${}^2$DG violates ground avoidance, GT and FP${}^2$DG show vertical landing at the last moment.

The overall results indicate that the GT method achieves fuel-efficient pinpoint landings while satisfying glide slope constraints and maintaining a terminal vertical landing. The GT method follows a concave trajectory under safety margin $C_\beta$ provides control margins for disturbances and model uncertainties. As a result, the GT method requires slightly more fuel compared to an optimal trajectory.

Lastly, the average CPU time usage for GT, ZEM/ZEV (fixed $t_f$), and FP${}^2$DG (fixed $t_f$) is $2.5\times 10^{-5}$ sec, $9.6\times 10^{-6}$ sec, and $9.7\times 10^{-6}$ sec, respectively. In the case of TZEM/ZEV (fixed $t_f$), it consumes $9.6\times 10^{-6}$ sec for Scenarios 1 and 2, and $4.5\times 10^{-5}$ sec for Scenario 3. This suggests that all methods are well suited for onboard implementation. For benchmark reference (though not a fair comparison), the second-order cone programming \cite{Acikmese_2007} with fixed $t_f$ and $N = 78$ took an average of $0.382$ sec for Scenario 3, and GPOPS took approximately $46$ sec to find the optimal solution. The test was conducted using the following configuration: CPU: I7-9700, RAM: 32 GB, Environment: MATLAB 2021a.

\subsection{Performance Analysis under Different Initial Downrange} \label{sec5.2}
\begin{figure}[!t]
\centering
\includegraphics[width=.45\textwidth]{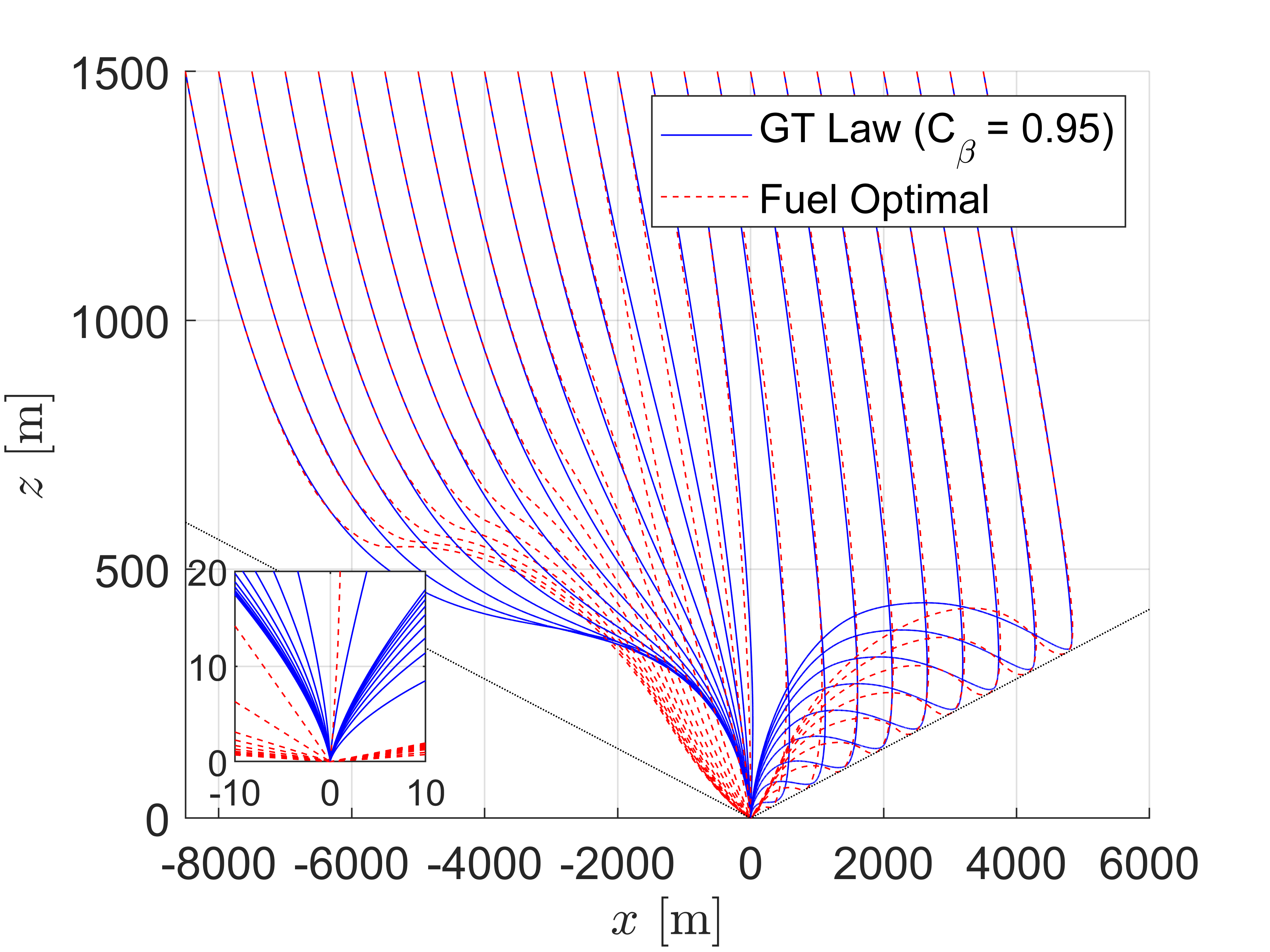} 
\includegraphics[width=.45\textwidth]{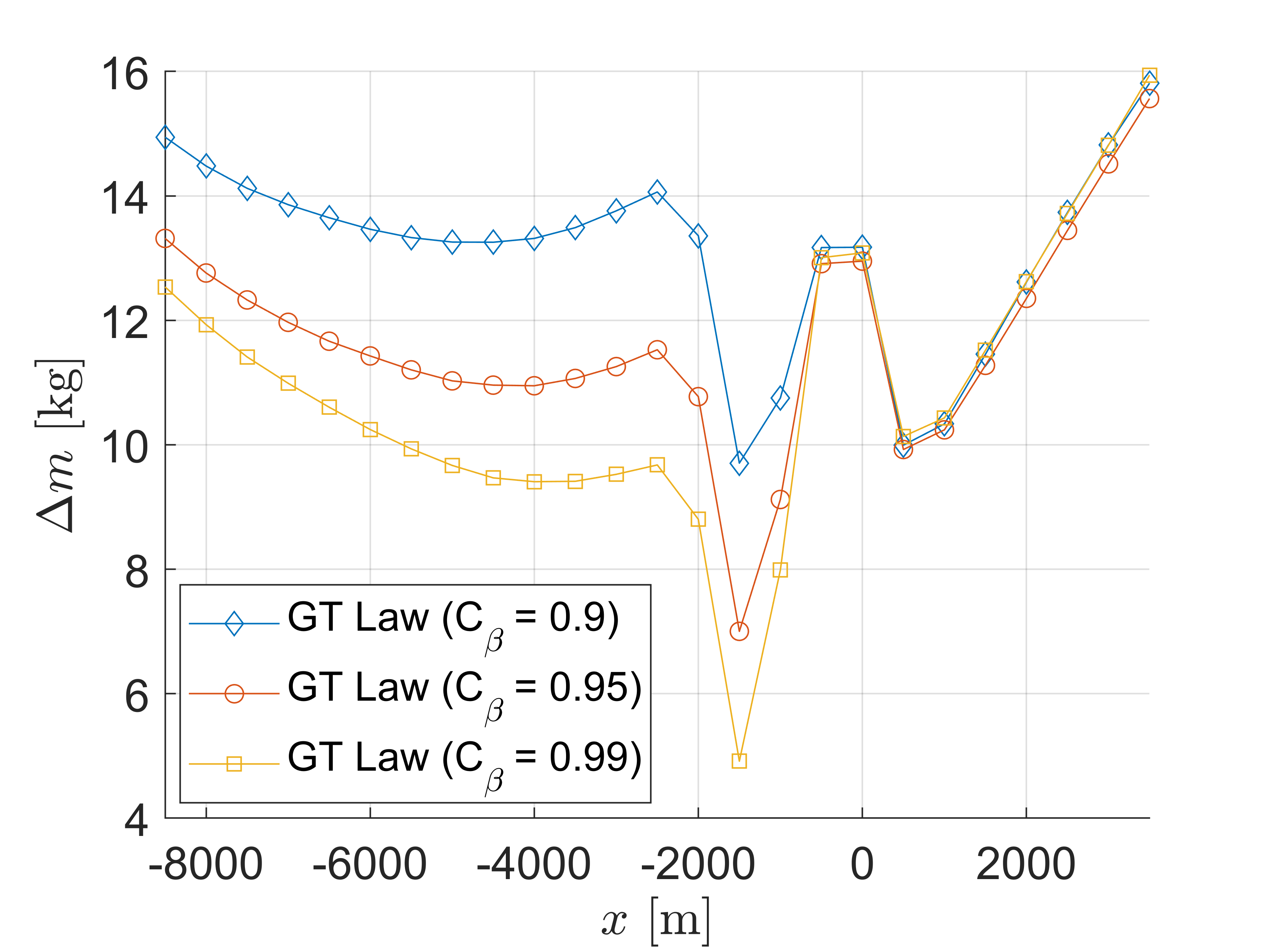}
\caption{Performance Comparison Results with glide slope constraint (a) Trajectory (b) Fuel Usage}
\label{fig_perform1}
\end{figure}

\begin{figure}[!t]
\centering
\includegraphics[width=.45\textwidth]{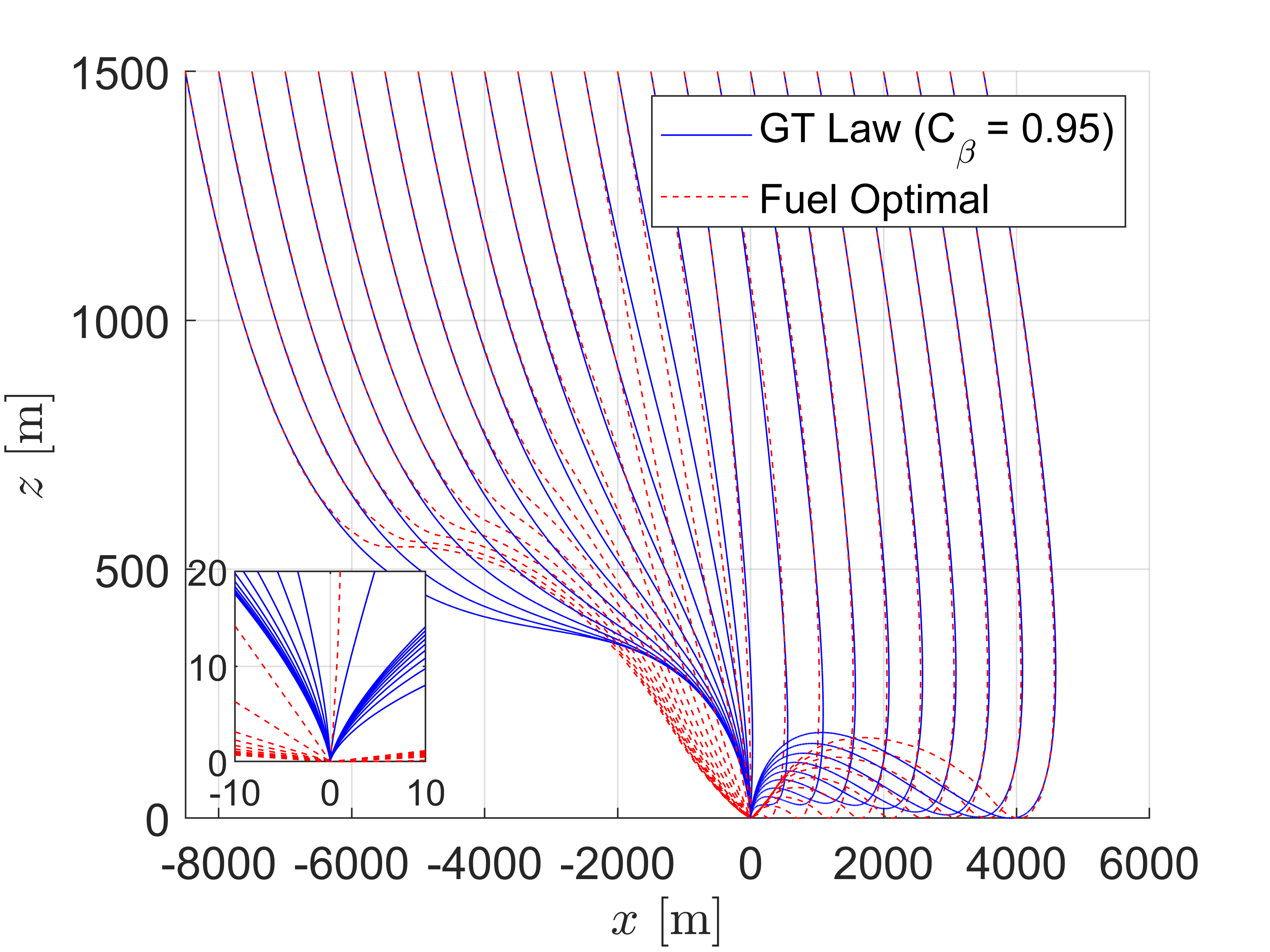} 
\includegraphics[width=.45\textwidth]{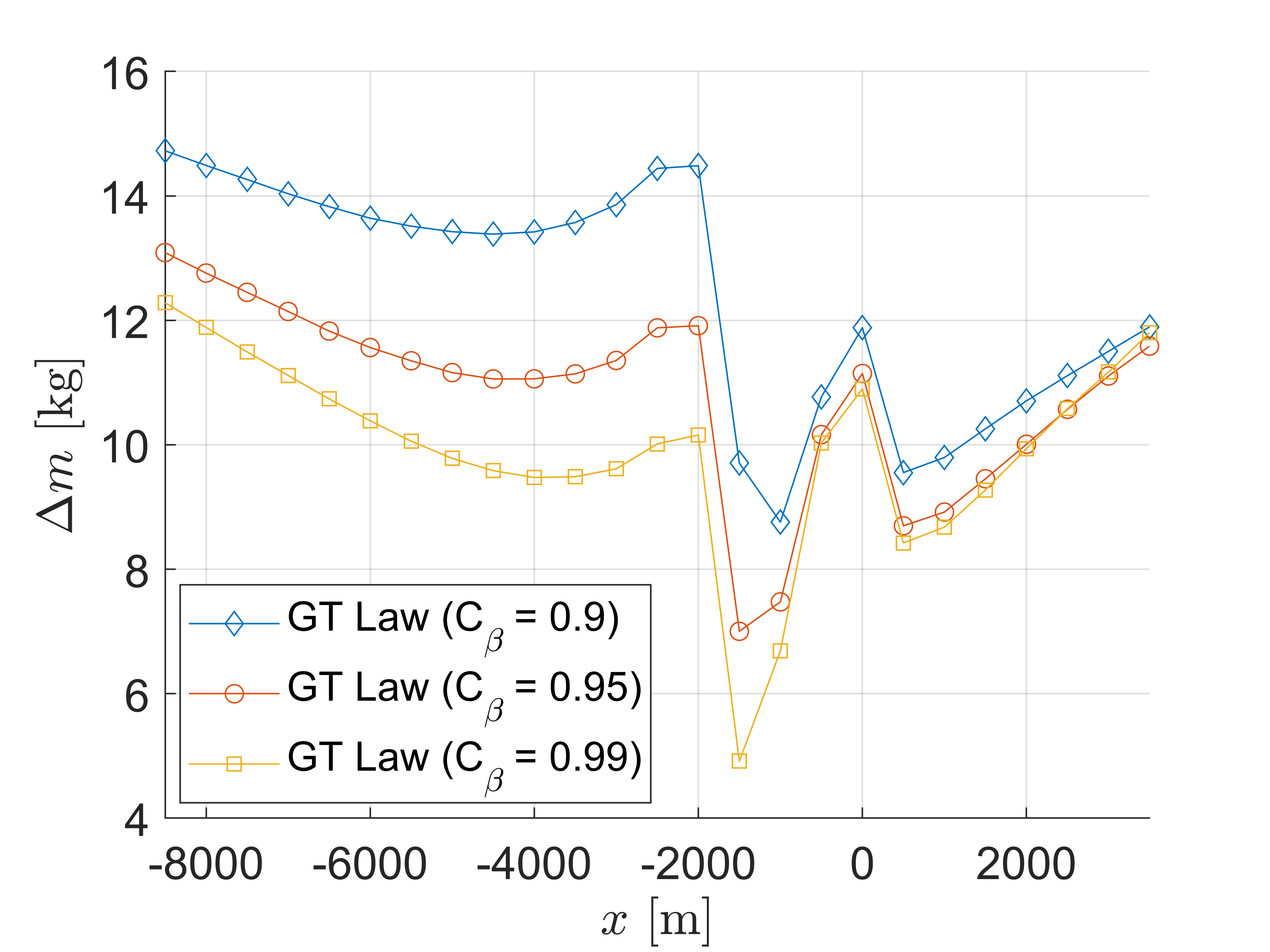}
\caption{Performance Comparison Results without glide slope constraint (a) Trajectory (b) Fuel Usage}
\label{fig_perform2}
\end{figure}

To further analyze the performance of the proposed algorithm, Scenario 3 in Sec. \ref{sec5.1} is tested with a wide range of initial downrange values, as conducted in prior studies \cite{ZHANG_smc, Wang_twophase}. Trajectories with and without a 4-degree glide slope constraint are illustrated in Fig. \ref{fig_perform1}.a and Fig. \ref{fig_perform2}.a, respectively. The additional fuel usage relative to the fuel optimal for each case is depicted in Fig. \ref{fig_perform1}.b and Fig. \ref{fig_perform2}.b. Additionally, the impact of different $C_\beta$ values on the performance is presented to provide insights into their effects. The fuel usage increases as the initial trajectory deviates from the GT trajectory, and reaches a minimum as the initial trajectory approaches the GT trajectory, which occurs around $x_0 = -1000$ m. Furthermore, as discussed in Section \ref{sec5.1}, the additional fuel usage can be primarily attributed to the choice of $C_\beta$ and the concavity of the trajectory. While higher values of $C_\beta$ tend to result in lower fuel usage, but it lead to greater sensitivity to disturbances and model uncertainties. Note that if the fuel-optimal trajectory considers the thrust constraint and thrust margin for the practicality, the difference in fuel usage becomes much smaller.

\subsection{Robustness Analysis Using Monte Carlo Simulation} \label{sec5.3}
While the proposed law has demonstrated promising performance based on the previous analysis, it is imperative to subject it to rigorous testing across various scenarios to ascertain its stability. Additionally, ensuring robustness against disturbances and model uncertainty is vital for practical implementation. To this end, Monte Carlo simulations are conducted to validate the proposed guidance law.

\begin{table}[!b]
\caption{\label{tab:robust_para} Parameters Distribution Used for Robustness Analysis}
\centering
\begin{tabular}{l c c}
\hline
Parameters    & Distributions \\\hline
$\textbf{\textit{r}}(t_0)$ & $\left[\mathcal{N}(500, 100^2),\ \mathcal{N}(500, 100^2),\ \mathcal{N}(1500, 100^2) \right]^\top$ m\\
$\textbf{\textit{v}}(t_0)$  & $\left[\mathcal{N}(100, 10^2),\ \mathcal{N}(10, 5^2),\ \mathcal{N}(-75, 5^2) \right]^\top$ m/s\\
$\eta$  &  $\mathcal{U}(-0.04, 0.04)$\ \%\\
$\xi$  &  $\mathcal{N}(0, 0.003)$\ \% \\
$\mu_i$  &  $\mathcal{U}(-0.3, 0.3)$ deg  for $i=1,2,3$\\
$\lambda_i$ & $\mathcal{U}(-0.02, 0.02)$  for $i=x,y,z$\\
\hline
\end{tabular}
\end{table}

Abnormal initial conditions with state dispersion defined in Table \ref{tab:robust_para} are used for the analysis where $\mathcal{N}(a,b)$ is a normal distribution and $\mathcal{U}(a,b)$ is a uniform distribution. Furthermore, the thrust and disturbance models employed in the simulation are as follows:
\begin{equation}
\begin{aligned}
    \textbf{\textit{T}} = m(1+\eta + \xi) M \textbf{\textit{u}}  , \quad \textbf{\textit{d}} = g\boldsymbol{\lambda} - \frac{1}{2}\rho C_D S_\text{ref} v^2 \hat{\textbf{\textit{v}}}
\end{aligned}
\end{equation}
where $\eta$ represents the thrust scale factor, $\xi$ represents the thrust instability factor, $M \equiv R_1(\mu_1)R_2(\mu_2)R_3(\mu_3)$ denotes the Euler 3-2-1 rotation misalignment matrix, $\boldsymbol{\lambda}^L \equiv \left[ \lambda_x, \lambda_y, \lambda_z \right]^\top$ is the bias disturbance vector, $\rho = 0.0274 \text{kg/m}^{3}$ is density of Mars atmosphere, $C_D = 1.0$ is drag coefficient, and $S_\text{ref}= 5\ \text{m}^2$ is the reference surface area of the lander. Each random parameter is sampled for every simulation under a uniform distribution governed by the parameters summarized in Table. \ref{tab:robust_para}.
In consideration of the worst-case combination of parameters, the effective thrust can potentially decrease by up to 10$\%$, so a value of $C_\beta = 0.85$ is used to avoid the risk of thrust saturation at the last moment. Note that the drag effect, which aids in the landing process, is neglected for the purpose of testing worst-case robustness, and cases with infeasible glide slope constraints are excluded from the analysis.

\begin{figure}[!t]
\centering
\includegraphics[width=.49\textwidth]{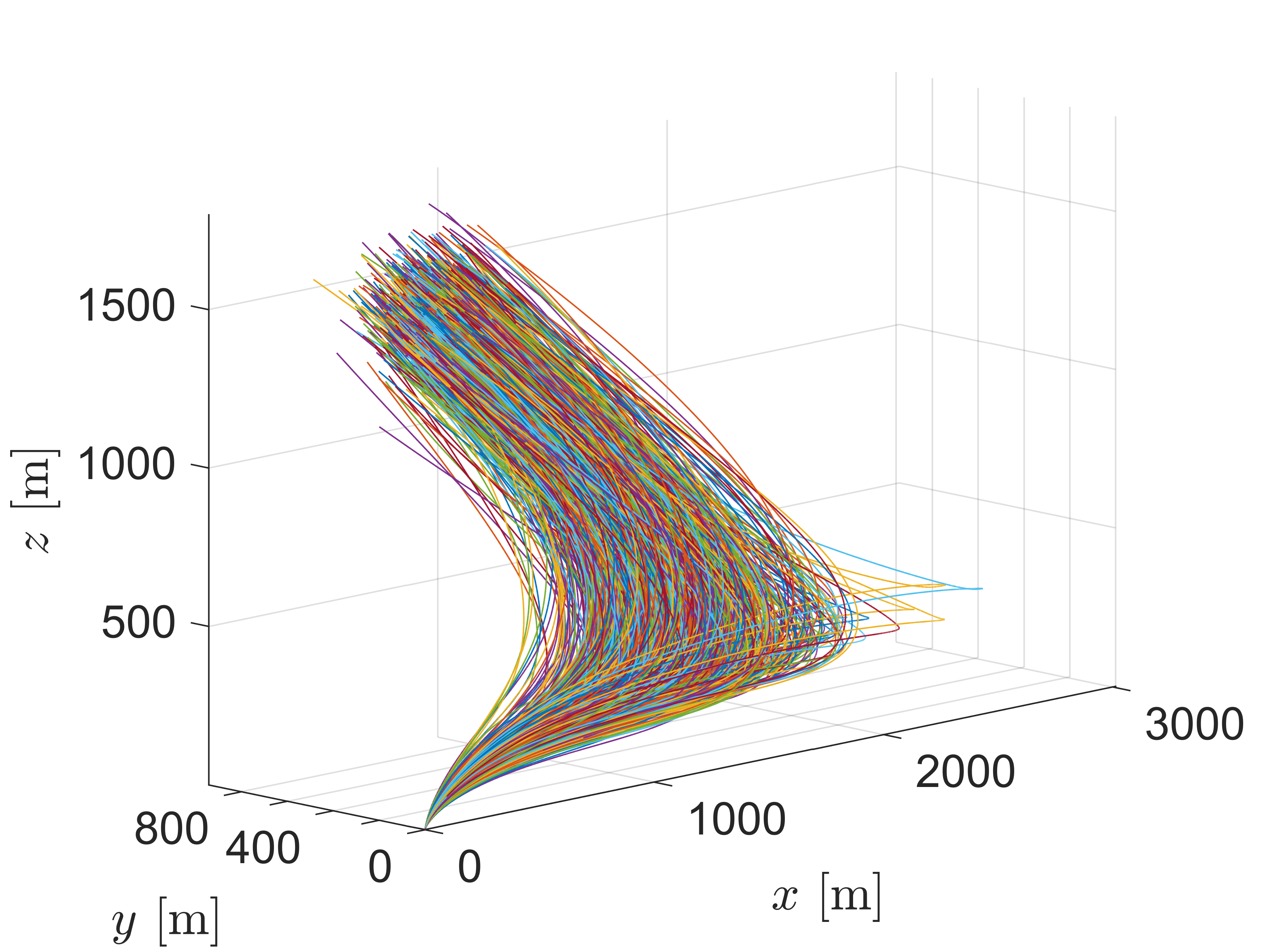} 
\caption{Vehicle trajectories under Monte Carlo simulations}
\label{fig_monte_traj}
\end{figure}

\begin{figure}[!t]
\centering
\includegraphics[width=.49\textwidth]{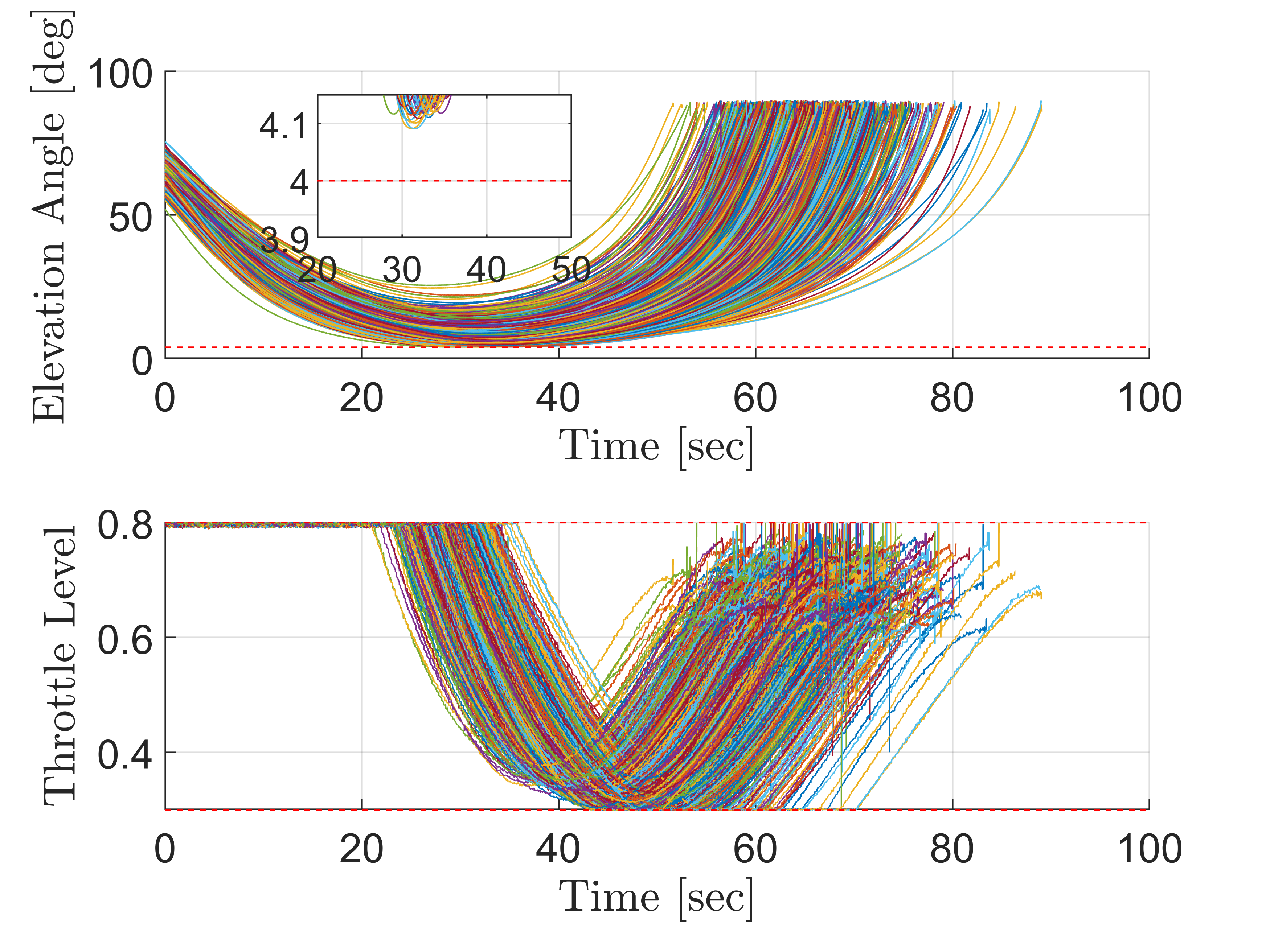}
\caption{Vehicle elevation angle and throttle level histories under Monte Carlo simulations}
\label{fig_monte_elv_lvl}
\end{figure}

\begin{figure}[!t]
\centering
\includegraphics[width=.49\textwidth]{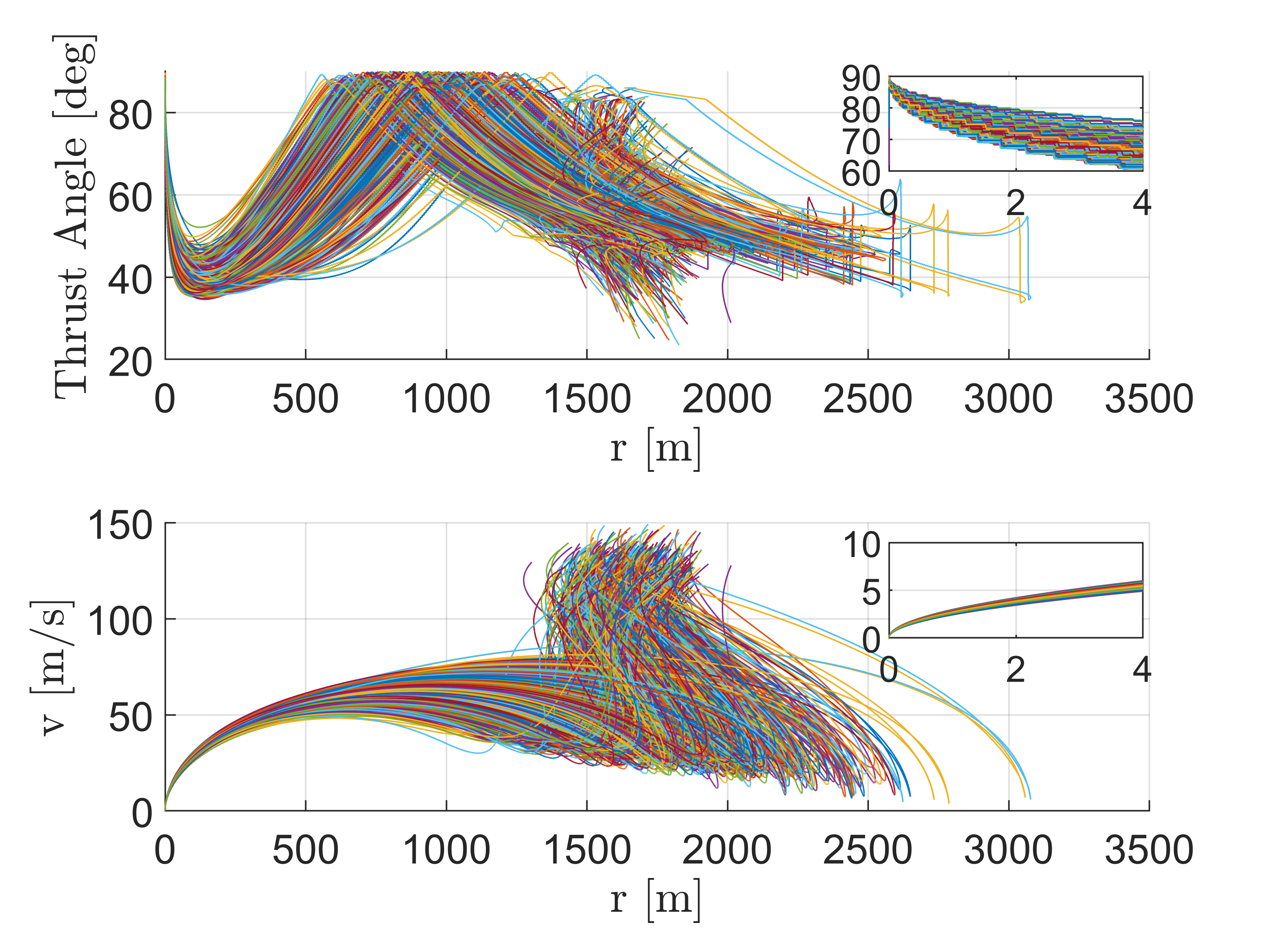}
\caption{Vehicle thrust angle and speed versus range under Monte Carlo simulations}
\label{fig_monte_pitch_err}
\end{figure}

Figure.\ref{fig_monte_traj} shows the trajectories of 1000 simulation samples, and Fig.\ref{fig_monte_elv_lvl} shows the elevation angle of $\textbf{\textit{r}}$ with respect to the frame $L$ and throttle level. As shown in the elevation angle graph, the vehicle satisfied the 4 degrees glide slope constraint and landed vertically even with unknown disturbances. Furthermore, Fig. \ref{fig_monte_pitch_err} illustrates the $\theta_u$, along with a graph depicting the range $r$ versus the speed $v$. The nearly 90-degree thrust angle at $r=0$ confirms that the thrust vector is aligned vertically at the moment of touchdown. Lastly, the last graph verifies that the proposed method satisfies the termination condition of $r < 0.01$ m and $v < 0.05$ m/s for all simulation cases. These results demonstrate the robustness and effectiveness of the proposed method in achieving precise and controlled landings under uncertain conditions.

\section{Conclusion} \label{conclusion}
This paper presents a novel guidance method for the terminal phase of EDL, utilizing a velocity vector generated by the gravity turn trajectory. The research investigates the advantageous properties of the gravity turn trajectory in the context of pinpoint landing and leverages them to design a guidance law. The proposed guidance law enables the lander to navigate safely around ground obstacles and achieve a desired vertical landing attitude at the last moment. Comparative evaluations against various feedback guidance laws demonstrate the effectiveness of the proposed approach in achieving pinpoint landings across a wide range of initial conditions while satisfying the glide slope constraint. Moreover, the proposed method achieves fuel consumption comparable to optimal solutions, despite achieving a vertical landing trajectory and incorporating an additional thrust margin for enhanced robustness. Monte Carlo simulations further validate the practicality and reliability of the proposed guidance approach for real-world applications.

\section*{Appendix: Proof of Eq.~\eqref{eq_law1}}
Firstly, we will show that $\Omega \textbf{\textit{e}}^G$ and $\boldsymbol{\omega}^G_{G/L} \times \textbf{\textit{v}}_d^G$ are identical:
\begin{equation} \label{eq_convert1}
\begin{aligned}
    \boldsymbol{\omega}^G_{G/L} \times \textbf{\textit{v}}_d^G = \begin{bmatrix}
        0 \\ \frac{v_x^\ast v_{y_G}}{x_\text{go}} \\ 0
    \end{bmatrix} = 
    \begin{bmatrix}
        0 \cdot (v_x^\ast - v_{x_G})\\ \frac{v_x^\ast}{x_\text{go}} \cdot (0 - v_{y_G}) \\ 0 \cdot (v_z^\ast - v_{z_G})
    \end{bmatrix} = \Omega \textbf{\textit{e}}^G
\end{aligned}
\end{equation}
Next, explicitly computing $F_{r_\text{go}} \textbf{\textit{v}}^G  +  F_{r_\text{go}}\textbf{\textit{e}}^G$ gives:
\begin{equation} \label{eq_convert2}
\begin{aligned}
    F_{r_\text{go}} \textbf{\textit{v}}^G + F_{r_\text{go}}\textbf{\textit{e}}^G = 
    F_{r_\text{go}} \textbf{\textit{v}}_d^G = \begin{bmatrix}
        -(4\beta^2 -1) g v_x^\ast \\ 0 \\ -(4\beta^2 - 4) g v_z^\ast
    \end{bmatrix}
\end{aligned}
\end{equation}
and expanding $F_{v_d} \left(  - \frac{\beta g}{v_d} \textbf{\textit{v}}^G_d+ \textbf{\textit{g}}^G \right)$ yields
\begin{equation} \label{eq_convert3}
\begin{aligned}
     F_{v_d} \left( - \frac{\beta g}{v_d} \textbf{\textit{v}}^G_d+ \textbf{\textit{g}}^G \right) &= \begin{bmatrix} 
    \frac{\partial f_x}{\partial v_x^\ast}
    & 0 & \frac{\partial f_x}{\partial v_z^\ast} \\ 0 & 0 & 0 \\ \frac{\partial f_z}{\partial v_x^\ast} & 0 & \frac{\partial f_z}{\partial v_z^\ast}
\end{bmatrix}
\begin{bmatrix} -\beta g \frac{v_x^\ast}{v_d} \\ 0 \\ -\beta g \frac{v_z^\ast}{v_d} - g \end{bmatrix}\\
&= -\frac{g}{v_d^2}\begin{bmatrix}
    2\beta^2 v_x^\ast{}^3 + 2\beta v_d^2 v_x^\ast + 2\beta v_x^\ast v_z^\ast{}^2  
    - v_d^2 v_x ^\ast  
    \\ 0 \\ 2\beta^2 v_x^\ast{}^2 v_z^\ast + 2\beta^2 v_z^\ast{}^3 + 2\beta^2 v_d^2 v_z^\ast - 4 v_d^2 v_z^\ast
\end{bmatrix} = \begin{bmatrix}
        -(4\beta^2 -1) g v_x^\ast \\ 0 \\ -(4\beta^2 - 4) g v_z^\ast
    \end{bmatrix}
\end{aligned} 
\end{equation}
Therefore, $F_{r_\text{go}} \textbf{\textit{v}}^G  +  F_{r_\text{go}}\textbf{\textit{e}}^G = F_{v_d} \left(  - \frac{\beta g}{v_d} \textbf{\textit{v}}^G_d+ \textbf{\textit{g}}^G \right) $ and this implies $ F_{v_d}^{\dagger}F_{r_\text{go}} \textbf{\textit{v}}^G  - \textbf{\textit{g}}^G = - \frac{\beta g}{v_d} \textbf{\textit{v}}^G_d  -F_{v_d}^{\dagger}F_{r_\text{go}}\textbf{\textit{e}}^G$, which proves that the first and second line of Eq.~\eqref{eq_law1} are equivalent.

\bibliography{references}   

\end{document}